\documentclass[letter,12pt]{article}

\usepackage[english]{babel}
\usepackage{booktabs}
\usepackage{rotating}
\usepackage{threeparttable, tablefootnote}
\usepackage{multirow}
\usepackage{color}
\usepackage{xcolor}
\usepackage{enumitem}
\usepackage{graphicx}
\usepackage{url} 

\usepackage{changepage}   
\usepackage[margin=1in]{geometry}
\usepackage{amsfonts}
\usepackage{amsmath}
\usepackage{amssymb}
\usepackage{bm}
\usepackage{dsfont}
\usepackage{xfrac}
\usepackage{mathtools}

\usepackage{floatrow}   
\usepackage{subcaption}

\usepackage{subfig}

\usepackage{natbib}
\newcommand{\mb}[1]{\mathbf{#1}}                
\renewcommand{\b}[1]{\mathbf{#1}}               
\newcommand{\norm}[1]{\left\lVert#1\right\rVert} 
\newcommand{\normsimple}[1]{\lVert#1\rVert}      

\newcommand{\mc}[1]{\mathcal{#1}}
\newcommand{\normF}[1]{\norm{#1}_F^2}
\newcommand{\normQuad}[1]{\norm{#1}_2^2}
\newcommand{\bigO}{\mathcal{O}}

\DeclareMathOperator*{\argmin}{arg\,min}

\usepackage{algorithm2e}
\SetKwComment{Comment}{/* }{ */}
\RestyleAlgo{ruled}

\usepackage{authblk}

\renewcommand{\b}[1]{\mathbf{#1}}               

\usepackage{mathtools}

\newcommand{\normOff}[1]{\norm{#1}_1^{\text{off}}}
\newcommand{\logl}[1]{\ell_{#1}(\mb{\Theta}_{#1})}

\newcommand{\T}{\b{\Theta}}
\newcommand{\Tcvn}{\b{\Theta}_{\text{CVN}}}
\newcommand{\Tcvnest}{\widehat{\b{\Theta}}_{\text{CVN}}}
\newcommand{\Tall}{\{\b{\Theta}_i\}_{i = 1}^m}

\newcommand{\Z}{\b{Z}}

\newcommand{\Y}{\b{Y}}

\newcommand{\wt}[1]{\widetilde{#1}}
\newcommand{\wh}[1]{\widehat{#1}}
\newcommand{\R}{\mathds{R}}

\newcommand{\Cov}{\b{\Sigma}}
\newcommand{\Lagr}[1]{\mathcal{L}_\rho (#1)}

\newcommand{\sumim}{\sum_{i = 1}^m}
\newcommand{\sumij}{\sum_{i < j}}

\newcommand{\indicator}[1]{\mathds{1}(#1)}

\begin{document}

\title{Inferring High-Dimensional Dynamic Networks Changing with Multiple Covariates}

\author[1]{Louis Dijkstra}
\author[1]{Arne Godt}
\author[1]{Ronja Foraita\footnote{Corresponding author. E-mail: \texttt{foraita@leibniz-bips.de}}}

\affil[1]{Leibniz Institute for Prevention Research \& Epidemiology -- BIPS, Bremen, Germany}


\maketitle

\begin{abstract}
\noindent 
\noindent High-dimensional networks play a key role in understanding complex relationships. These relationships are often dynamic in nature and can change with multiple external factors (e.g., time and groups). Methods for estimating graphical models are often restricted to static graphs or graphs that can change with a single covariate (e.g., time). We propose a novel class of graphical models, the covariate-varying network (CVN), that can change with multiple external covariates.

In order to introduce sparsity, we apply a $L_1$-penalty to the precision matrices of $m \geq 2$ graphs we want to estimate. These graphs often show a level of similarity. In order to model this `smoothness', we introduce the concept of a `meta-graph' where each node in the meta-graph corresponds to an individual graph in the CVN. The (weighted) adjacency matrix of the meta-graph represents the strength with which similarity is enforced between the $m$ graphs.

The resulting optimization problem is solved by employing an alternating direction method of multipliers. We test our method using a simulation study and we show its applicability by applying it to a real-world data set, the gene expression networks from the study `German Cancer in childhood and molecular-epidemiology' (KiKme). An implementation of the algorithm in R is publicly available under \url{github.com/bips-hb/cvn}. \\[5pt]
\end{abstract}

\section{Introduction}
\label{sec:intro}

In many scientific fields, network models are essential for describing and understanding complex relationships \citep{newman2018networks}. In genetic and molecular epidemiology, networks elucidate interactions between molecular regulators governing gene expression such as gene regulatory networks  \citep{barabasi2004network} and how microbial communities in the microbiome interact with each other and the host \citep[e.g.][]{huttenhower2012, deVos2012}. In neuroscience, network models map intricate brain connections, aiding in understanding communication and coordination between different regions \citep{badhwar2017resting}.

Extensive research has been devoted to inferring high-dimensional networks from data \citep{Friedman2008,Cai2011,Danaher2014}, primarily on static networks where edges remain fixed and unaffected by external influences. In this paper,  we use the terms `graph' and `network' interchangeably. While these static networks provide valuable insights, it is crucial to also understanding how network structures evolve with respect to external covariates, such as time or group differences.
For example, long-term exposure to ultraviolet (UV) radiation can alter gene networks, impairing DNA repair mechanisms and potentially leading to skin cancer \citep{brozyna2007mechanism}. Microbiome networks differ between healthy individuals and those with obesity \citep{de2018gut}. In neuroepidemiology, changes in brain connectome interactions due to cholesterol levels and aging indicate Alzheimer’s disease 
\citep{badhwar2017resting}. Observing how dynamic networks change with external factors -- such as UV radiation, treatment regimes, or cholesterol levels -- is fundamental to understanding the underlying biological processes.

Current literature provides, to be best of our knowledge, only methods for analyzing networks that change with a single discrete external covariate, such as time \citep{Monti2014,Hallac2017,gibberd2017regularized} or groups \citep{Danaher2014,Hallac2017,vinciotti2024random}. Here, we propose extending these methods to multiple, possibly intricately related, discrete covariates. 
We present a very general framework for defining a dynamical graphical model of this kind. We call this the \emph{covariate-varying network} (CVN) model. We present an alternating direction method of multipliers (ADMM) algorithm that can be used to estimate the model on the basis of (potentially high-dimensional) data. 

The paper is organized as follows: Section~\ref{sec:staticnetworks} introduces static Gaussian graphical models in which the (in)dependency structure is unaffected by external covariates; we simultaneously present much of the notation used throughout the paper. In Section~\ref{sec:cvn}, we formally define the CVN graphical model, including the concept of a \emph{meta-graph} to represent the similarities between the various individual graphs. This section also discusses the estimator considered throughout the paper. Additionally, we highlight several known models from the literature that are special cases of the CVN framework.

We solve the estimation problem for the CVN using an ADMM algorithm that we describe in detail in Section~\ref{sec:algorithm}. The computational complexity of the algorithm is determined in Section~\ref{sec:computationalComplexity}. We show how one could interpolate a graph given an estimated CVN in Section~\ref{sec:interpolation}. The two tuning parameters of the model, one regulating sparsity and the other controling the level of similarity between the graphs, need to be chosen. In Section~\ref{sec:tuningParameterSelection}, we show how this could be done using either the Akaike or Bayesian information criterion (AIC and BIC). We propose an alternative tuning parameterization in Section~\ref{sec:alternativeTuningParameterization} that leads to more stable choices that are barely influenced by the introduction of more variables and/or covariates. 

We assess the performance of the method in a simulation study in which we explore to what extent the proposed method can reconstruct a CVN. We present the simulator in Section~\ref{sec:simulation} and the simulation set-up itself in Section~\ref{sec:simulationSetUp}. The measures used to assess the performance are presented in the latter section as well. 

In addition, we apply the method to a real data set from the KiKme study \citep{Marron2021}, see Section~\ref{sec:realDataAnalysis}. The results of both the simulation and this case study are shown in Section~\ref{sec:results}. We end with our conclusions and a discussion in Section~\ref{sec:conclusions}.

The ADMM algorithm has been implemented as an \texttt{R}~package and is publicly accessible at \url{www.github.com/bips-hb/CVN}. Additionally, the CVN data simulator is available as an \texttt{R} package at \url{www.github.com/bips-hb/CVNSim}, and the code for the simulation study can be found at \url{www.github.com/bips-hb/CVNStudy}. All results of the simulation study are hosted at \url{cvn.bips.eu}. 

\section{Static Networks} \label{sec:staticnetworks}

We represent a static network as a probabilistic undirected graphical model with the tuple $\{\b{X}, f(\b{X}), G = (V,E)\}$ where $\b{X} = (X_1, X_2, \ldots, X_p)^\top$ is a real-valued $p$-dimensional random vector with joint density function $f(\b{X})$.  The nodes of the graph $G$, $V = \{1,2,\ldots, p\}$, correspond to the variables $X_1, X_2, \ldots, X_p$. The edges $E \subseteq V \times V$ reflect the (in)dependence structure between the variables in accordance with the density function $f(\b{X})$. The presence or absence of an edge represents whether the two corresponding variables are conditionally independent, i.e., the edge $\{s,t\}$ is in $E$ if $X_s \not\!\perp\!\!\!\perp X_t \mid \mb{X}_{V \setminus \{s,t\}}$, where $\mb{X}_{V \setminus \{s,t\}}$ are all variables in $\b{X}$ except $X_s$ and $X_t$. 
Let $\b{A} = (a_{st})_{p \times p}$ be the graph's adjacency matrix where $a_{st} = \indicator{\{s,t\} \in E}$ and $\indicator{\cdot}$ is the indicator function. 

We assume the random vector $\b{X}$ to follow a multivariate normal distribution with mean $\bm{\mu}$ and covariance matrix $\Cov$. Under this assumption, the entries of the precision matrix $\T = \left(\theta_{st} \right)_{p \times p} = \Cov^{-1}$ are proportional to the partial correlations between the variables. Determining the network structure, i.e., the edge set $E$, is, therefore, equivalent to determining which entries in the precision matrix $\T$ are zero and non-zero. In terms of the graph's adjacency matrix, $a_{st} = \indicator{\theta_{st} \neq 0}$ for all $\{s,t\} \in V \times V$ \citep[see][]{lauritzen1996}. Without loss of generality, we assume that the data are centered around the mean $\bm{\mu} = \b{0}$ for the remainder of the paper. 

Suppose we observe $n$ realizations $\bm{x}_1, \bm{x}_2, \ldots, \bm{x}_n$ of the random vector $\b{X}$. The maximum likelihood estimator of the precision matrix is given by
\begin{equation}
    \widehat{\T} = \argmin_{\T\succ 0} \logl{} = \argmin_{\T \succ 0}~ \frac{n}{2} \left[\text{trace}\left( \widehat{\Cov} \T \right) - \log \text{det}\left(\Cov \right) \right],
    \label{eq:orig}
\end{equation}
where $\T \succ 0$ denotes positive definiteness, $\logl{}$ is the log-likelihood function, `$\text{trace}$' and `$\text{det}$' denote the trace and determinant of a matrix, and $\widehat{\Cov} = n^{-1}\sum_{i=1}^n \bm{x}_i \bm{x}_i^\top$ is the empirical covariance matrix \citep[see, for example][]{Uhler2018}. However, this estimator does  lead to dense graphs. Moreover, in high-dimensional settings, the empirical covariance matrix $\widehat{\Cov}$ is rank deficient and cannot be used to estimate the precision matrix $\widehat{\T}$. One can estimate $\widehat{\T}$ by imposing a sparse graph structure and, hence, a sparse precision matrix.  
A popular method to estimate sparse networks is the Graphical Least Absolute Selection and Shrinkage Operator (GLASSO; \citet{Banjeree2008} and \citet{Friedman2008}) which places a $L_1$-norm penalty on the precision matrix in order to `shrink' its off-diagonal entries to zero. It involves solving the following convex optimization problem:
\begin{equation}
    \widehat{\mb{\Theta}}_{\text{GLASSO}} = \argmin_{\T \succ 0} \ell(\mb{\Theta}) + \lambda \normOff{\T}. 
    \label{eq:GLASSO}
\end{equation}
Here, $\lambda > 0$ is the tuning parameter and $\normOff{\T} = \sum_{s \neq t} |\theta_{st}|$ is the off-diagonal $L_1$-norm of the precision matrix. 

\section{The Covariate-Varying Network Model} \label{sec:cvn}

We represent a covariate-varying network (CVN) graphical model with the quintuple
\begin{equation}
    \text{CVN} = \{\b{X}, \b{U}, \mc{U}, f(\b{X} \mid \b{U}), \{G(u) = (V, E(u)) \}_{u \in \mc{U}} \},
    \label{eq:cvnmodel}
\end{equation}
where $\b{X} = (X_1, X_2 \ldots, X_p)^\top$ is a $p$-dimensional random vector and $\b{U} = (U_1, U_2, \ldots, U_q)^\top$ is a random vector representing $q$ external discrete covariates, i.e., variables that are not included in the graph with $(K_1,\ldots, K_q)^\top$ categories. The vector $\b{U}$ lies in the discrete space $\mc{U}$ with cardinality $m \leq \prod_{k=1}^q K_k$. The joint density function of $\b{X}$ conditioned on $\b{U}$ is $f(\b{X} \mid \b{U})$. The fifth element of the CVN is a set of $m$ graphs, one for each value $u$ in $\mc{U}$. The vertices of $G(u)$, $V = \{1, 2, \ldots, p\}$, correspond to the variables $X_1, X_2, \ldots, X_p$ and do not change with $\b{U}$. The (in)dependence structure between the variables is captured by the edge set $E(u)$ which can change with $\b{U}$. Specifically, the edge $\{s,t\}$ is in $E(u)$ if $X_s \not\!\perp\!\!\!\perp X_t \mid U = u\ \text{and } \mb{X}_{V \setminus \{s,t\}}$. We denote the adjacency matrix of graph $G(u)$ as $\b{A}(u) = (a_{st}(u))_{p \times p}$ where $a_{st}(u) = \indicator{\{s,t\} \in E(u)}$. In this paper, we are interested in estimating the graphs $\{G(u) = (V, E(u)) \}_{u \in \mc{U}}$. 

We assume throughout that $\b{X} \mid \b{U} = u$ follows a multivariate normal distribution with mean $\bm{\mu}(u) = \b{0}$ and covariance matrix $\Cov(u)$. Under the normality assumption, the entries of the precision matrices correspond with the edge sets of the different graphs. Let $\T(u) = \Cov(u)^{-1} = (\theta_{st}(u))_{p \times p}$ be the precision matrix, then the edge $\{s,t\}$ is in $E(u)$ if $\theta_{st}(u) \neq 0$. Hence, estimating the structure of the graphs $\{G(u)\}_{u \in \mathcal{U}}$, corresponds to determining the zero and non-zero entries in the precision matrices $\{\T(u)\}_{u \in \mathcal{U}}$.  

We index each element in $\mc{U}$ by assigning them a unique element from the index set $\mc{I} = \{1, 2, \ldots, m\}$. If $i \in \mc{I}$ is the index of $u \in \mc{U}$, we write $G_i = G(u)$, $E_i = E(u)$, $\Cov_i = \Cov(u)$, $\T_i = \T(u)$ and $\b{A}_i = \b{A}(u)$. The $(s,t)$ entry in the $i$-th precision matrix is written as $\theta_{st}^{(i)}$. The collection of all precision matrices is denoted by $\Tcvn = \Tall$.

For each value of $u \in \mc{U}$, we obtain $n_i > 0$ observations:
$\left(\bm{x}_1^{(i)}, \bm{x}_2^{(i)}, \ldots, \bm{x}_{n_i}^{(i)}\right)$ where $\bm{x}_j^{(i)} \in \R^p$ and  $j = 1, \ldots,  n_j$. 
Similarly to eq.~\eqref{eq:orig}, we can express the log-likelihood function for precision matrix $\T_i$ as 
\begin{equation}
    \ell_i(\T_i) = \frac{n_i}{2} \cdot \left[\text{trace}\left( \widehat{\Cov}_i \T_i \right) - \log \text{det}\left(\Cov_i \right) \right],
    \label{eq:likelihood}
\end{equation}
where $\widehat{\Cov}_i = n{_i}^{-1} \sum_{j = 1}^{n_i} \bm{x}_j^{(i)} (\bm{x}_j^{(i)})^\top$ is the empirical covariance matrix. 
Overall, the log-likelihood for the CVN model itself is simply the sum over all graphs, i.e., 
$\sum_{i = 1}^m \ell_i(\T_i)$.

To introduce sparsity, we apply, just as for the GLASSO in eq.~\eqref{eq:GLASSO}, the off-diagonal $L_1$-norm penalty for each individual graph:
\begin{equation*}
    \argmin_{\Tall} \sum_{i = 1}^m \logl{i} + \lambda_1 \sumim \normOff{\T_i},
\end{equation*}
where $\lambda_1 > 0$ is the GLASSO tuning parameter. Larger values of $\lambda_1$ increase the penalty and result in a sparser estimate of the precision matrices.

In many applications, we expect some similarities between the graphs. For example, two networks observed on two consecutive time points tend to have similar edge sets, while networks observed on far removed time points show less similarity. We refer to enforcing similarity between certain graphs as \emph{smoothing}. For introducing smoothness across the $m$ graphs, we first define a \emph{meta-graph}. Let $\mc{G} = (\mc{V}, \mc{E}, w)$ be a weighted, undirected graph where each node in the node set $\mc{V} = \{1,2,\ldots,m\}$ corresponds to the graphs $G_1, G_2, \ldots, G_m$, $\mc{E} \subseteq \mc{V} \times \mc{V}$, and $w: \mc{E} \rightarrow [0,1]$ is the weight function that assigns a value between $0$ and $1$ to each edge $e \in \mc{E}$. We denote the symmetric weighted adjacency matrix of the meta-graph by $\b{W} = (w_{ij})_{m \times m}$ where $w_{ij} = w_{ji} = w(\{i,j\})$. 

The weight matrix reflects some prior knowledge about the similarity of how the $m$ graphs might change with the external covariates. For a consistent structure estimation, we must impose an additional smoothness constraint on $\Tall$ to detect graph-wise changes in the underlying (in)dependence structure.
We define the CVN estimator of  $\Tcvn = \Tall$ as
\begin{equation}
    \Tcvnest =
        \argmin_{\Tall} \sum_{i = 1}^m \logl{i} + \lambda_1 \sumim \normOff{\T_i} + \lambda_2 \sumij w_{ij} \cdot \normOff{\T_i - \T_j},
    \label{eq:cvn}
\end{equation}
where $\lambda_2 \geq 0$ is the \emph{smoothing tuning parameter}. 
The term $w_{ij} \normOff{\T_i - \T_j}$ penalizes the differences between the edge sets of graph $G_i$ and graph $G_j$. The value $w_{ij}$ encourages a similar structure between graphs $G_i$ and $G_j$ whereas $\lambda_2$ regulates how strongly differences between all $\binom{m}{2}$ pairs of graphs are penalized.

\subsection{Special Cases of Covariate-Varying Networks}\label{sec:specialcases}
By choosing the space $\mc{U}$, the indexation $\mc{I}$ and the weighted adjacency matrix $\b{W}$ of the meta-graph appropriately, we can express various models from the literature as special cases of the CVN graphical model. 

\begin{figure}
    \centering
    \includegraphics[width=\textwidth]{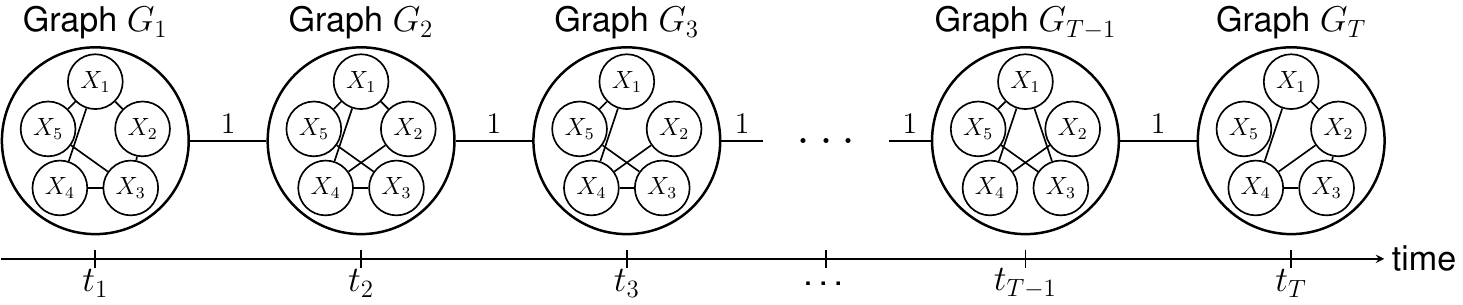}
    \caption{The Time-Varying Graphical LASSO (TVGL) as a special case of the CVN model, see eq.~\eqref{eq:weightmatrixTVGL} and \eqref{eq:TVGL}. A total of $T$ graphs $G_i$ with $p=5$ variables  $X_1, X_2, \ldots, X_5$ are observed at time points $t_1, t_2, \ldots, t_T$. 
    The circle around each graph represents a node in the meta-graph $\mc{G}_\text{TVGL}$; the links between the graphs are the meta-graph's edges. Graphs on consecutive time points are smoothed with weight $1$.}
    \label{fig:specialcaseTVGL}
\end{figure}

\paragraph{Time-Varying Graphical LASSO (TVGL)} Suppose we observe graphs at $T$ time points, i.e., $\mc{U} = \{t_1, t_2, \ldots, T_T\}$. We define the index set $\mc{I} = \{1,2, \ldots, T\}$ where $i \in \mc{I}$ corresponds to time point $t_i$. Consequently, the meta-graph $\mc{G}_{\text{TVGL}}$ has $T$ nodes, each corresponding to a single graph, and only graphs on consecutive time-points are smoothed. We define the $(T \times T)$-dimensional adjacency matrix of $\mc{G}_{\text{TVGL}}$ as  
\begin{equation}
   \b{W}_{\text{TVGL}} = \left(w_{ij}^{\text{TVGL}}\right)_{T \times T} \qquad \text{where } w_{ij}^\text{TVGL} = \begin{cases}
        1 & \text{if } |i - j| = 1 \\ 
        0 & \text{otherwise}. 
    \end{cases}
    \label{eq:weightmatrixTVGL}
\end{equation}
The time-varying graphical LASSO $\{\widehat{\T}_t^{\text{TVGL}} \}_{t = 1}^T$ can then be estimated by solving
\begin{equation}
    \{\widehat{\T}_t^{\text{TVGL}} \}_{t = 1}^T = \argmin_{\{\T_t \}_{t = 1}^T} \sum_{t = 1}^T \ell_t(\T_t) + \lambda_1 \sum_{t = 1}^T \normOff{\T_t} + \lambda_2 \sum_{t = 2}^T \normOff{\T_{t} - \T_{t - 1}}
    \label{eq:TVGL}
\end{equation}
which is equivalent to the definition in~\citet{Hallac2017} and \citet{Monti2014}. Figure~\ref{fig:specialcaseTVGL} shows an example of a TVGL in which each of the $T$ graphs has five nodes corresponding to $X_1, X_2, \ldots, X_5$.


\paragraph{Fused Graphical LASSO (FGL)} 

\begin{figure}[h!]
    \centering
    \includegraphics[width=.8\textwidth]{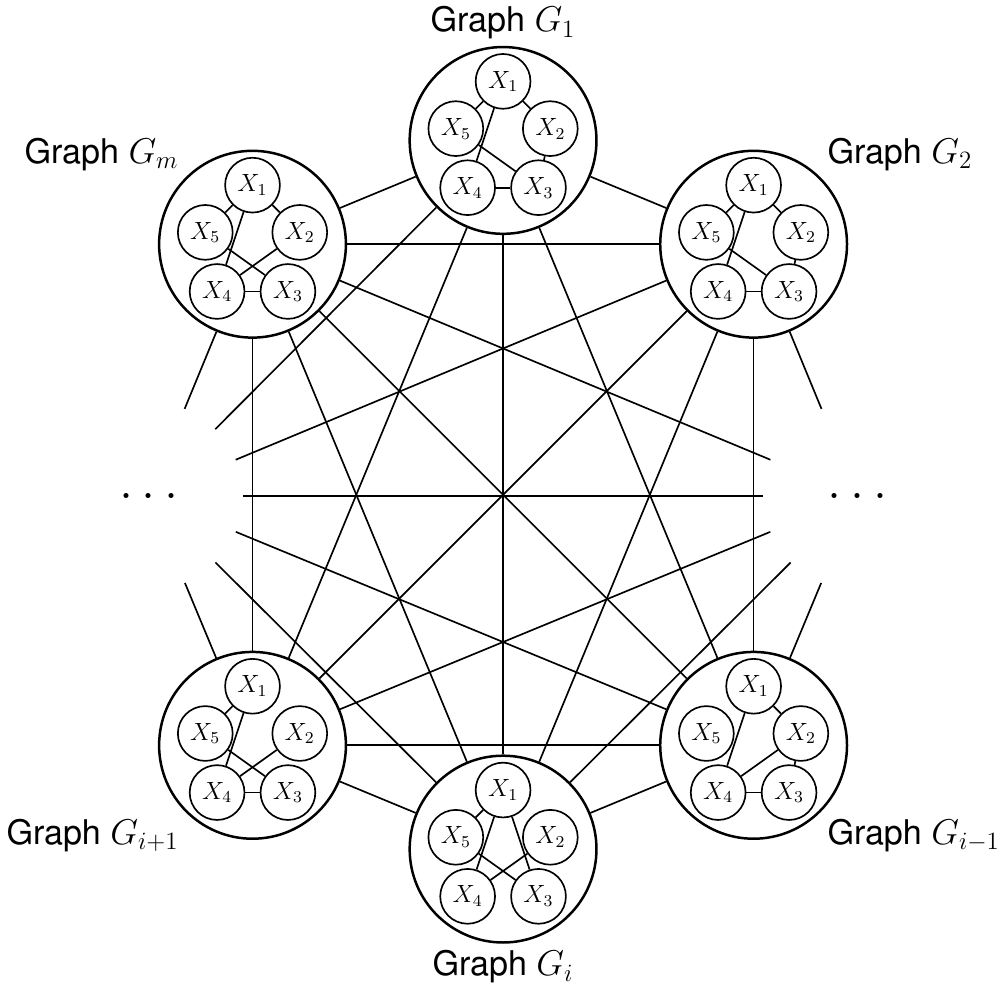}
    \caption{The joint graphical LASSO with the fused graphical LASSO (FGL) penalty term as a special case of the CVN model, see eq.~\eqref{eq:weightmatrixFGL} and \eqref{eq:FGL}. A total of $m$ graphs with $p=5$ variables are observed. Just as in Figure~\ref{fig:specialcaseTVGL}, 
    each graph represents a node in the meta-graph $\mc{G}_{\text{FGL}}$ and links between the graphs are the meta-graph’s edges.}
    \label{fig:specialcaseFGL}
\end{figure}

When there is only one external covariate,  the CVN reduces in some cases to the fused graphical LASSO (FGL) from \citet{Danaher2014}. The FGL takes advantage of structural similarities across different groups to estimate $m$ graphical models that share common edges.  Let $m$ be the number of categories of the single external variable $U$, and for each category there are $n_1, n_2, \ldots, n_m > 0$ observations. 
As before, we choose the index set $\mc{I} = \{1,2,\ldots,m\}$ where the graph of $i$-th category is denoted with $G_i$. The CVN reduces to the FGL with the fused graphical LASSO penalty term if the $(m \times m)$-dimensional weight matrix is defined as 
\begin{equation}
    \b{W}_{\text{FGL}} = (w_{ij}^{\text{FGL}})_{m \times m} \qquad \text{where } w_{ij}^\text{FGL} = \begin{cases}
        1 & \text{if } i \neq j \\ 
        0 & \text{otherwise}. 
    \end{cases}
    \label{eq:weightmatrixFGL}
\end{equation}
This leads to the following minimization problem:
\begin{equation}
    \{\widehat{\T}^{\text{FGL}}_i \}_{i = 1}^m  = \argmin_{\Tall} \sum_{i = 1}^m \ell_i(\T_i) + \lambda_1 \sum_{i = 1}^m \normOff{\T_i} + \lambda_2 \sum_{i < j} \normOff{\T_{i} - \T_{j}},
    \label{eq:FGL}
\end{equation}
which is the FGL \citep{Danaher2014}. 
Expressed as CVN this means that the meta-graph is fully connected with weights of $1$. The FGL penalty enforces that every graph is smoothed with every other graph.
See Figure~\ref{fig:specialcaseFGL} for a visual representation of this model.

\begin{figure}[h!]
    \centering
    \includegraphics[width=\textwidth]{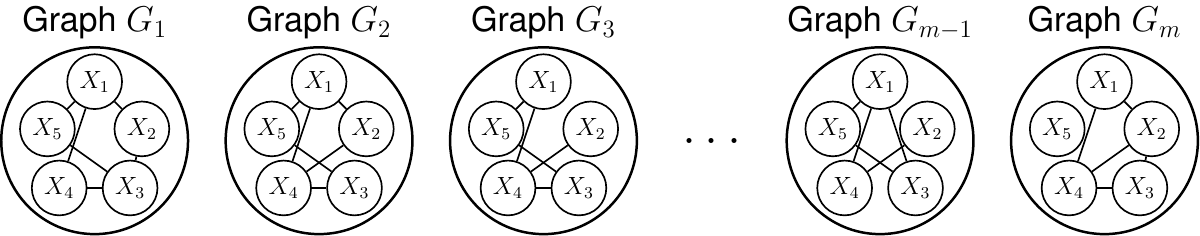}
    \caption{A trivial case with $\b{W}_0 = \b{0}$ without smoothing the $m$ graphs. 
    }
    \label{fig:staticnetworks}
\end{figure}

\paragraph{Multiple Static Graphical Models} A trivial case is when no structural constraints are imposed on the $m$ graphs. The weight matrix is hence defined as $\b{W}_0 = \b{0}$. The CVN reduces then to $m$ independent GLASSO problems as in eq.~\eqref{eq:GLASSO}: 
\begin{equation*}
    \{\widehat{\T}_i^0 \}_{i = 1}^m  = \argmin_{\Tall} \sum_{i = 1}^m \ell_i(\T_i) + \lambda_1 \sum_{i = 1}^m \normOff{\T_i} = \left\{ \argmin_{\T_i} \ell_i(\T_i) + \lambda_1 \normOff{\T_i} \right\}_{i = 1}^m.
    \label{eq:staticnetworks}
\end{equation*}
Figure~\ref{fig:staticnetworks} shows a visual representation.


\section{An Alternating Direction Method of Multipliers Algorithm for the Covariate-Varying Network Problem} \label{sec:algorithm}

Estimating the CVN graphical model defined in eq.~\eqref{eq:cvnmodel} requires us to solve the optimization problem in eq.~\eqref{eq:cvn}. We propose to do this using an Alternating Direction Method of Multipliers (ADMM) algorithm \citep{Boyd2004}. By doing so, we can split the original problem into three subproblems, which in turn can be solved either analytically or numerically.

First, let us introduce a set of \emph{consensus variables} $\Z = \{\Z_i\}_{i = 1}^m$ where $\Z_i = \T_i$ for all $i$. In addition, we write $\T = \{\T_i\}_{i = 1}^m$. We can then express eq.~\eqref{eq:cvn} as the equality-constrained convex optimization problem:
\begin{equation}
    \begin{gathered}
\argmin_{\T, \Z} \sum_{i=1}^m \logl{i} + \lambda_1 \sum_{i = 1}^m \normOff{\Z_i} + \lambda_2 \sumij w_{ij} \normOff{\Z_i - \Z_j} \\[1ex]
\text{subject to } \mb{\Theta}_i - \mb{Z}_i = \mathbf{0} \text{ and } \mb{\Theta}_i, \mb{Z}_i  \succ 0 \text { for }i = 1,2, \ldots, m. 
\end{gathered}
\label{eq:consensus}
\end{equation}
Thus, the problem does not have to be solved jointly but can be solved for $\Z$ and $\T$ separately. 
The corresponding augmented Lagrangian is 
\begin{equation}
    \begin{split}
    \Lagr{\T, \Z, \Y} & = \sum_{i = 1}^m \logl{i} + \lambda_1 \sum_{i = 1}^m \normOff{\Z_i} + \lambda_2 \sumij w_{ij} \normOff{\Z_i - \Z_j} \\ 
    & + (\rho/2) \sum_{i = 1}^m \left[ \normF{\T_i - \Z_i + \Y_i} - \normF{\Y_i} \right],
    \end{split}
    \label{eq:lagrangian}
\end{equation}
where $\Y = \{\Y_i\}_{i=1}^m$ is a set of \emph{scaled dual variables}, $\normF{\b{A}} = \left(\sum_{s,t} a_{st}^2\right)^{1/2}$ is the Frobenius norm, and $\rho > 0$ is the ADMM's \emph{dual update step length} \citep{Boyd2004, Danaher2014, Hallac2017}.
Minimizing the function in eq.~\eqref{eq:lagrangian} 
is equivalent to minimizing the original problem in eq.~\eqref{eq:cvn}. The advantage, however, of this alternative definition, is that eq.~\eqref{eq:consensus} can now be solved by iteratively updating $\T$, $\Z$ and $\Y$ \citep{Boyd2004}. The update steps for the ADMM are 
\begin{align}
     \text{Likelihood update: } &   \T(k + 1) = \argmin_{\T}~ \Lagr{\T, \Z(k), \Y(k)}, \label{eq:updateTheta} \\ 
     \text{Constraint update: } &
        \Z(k + 1) = \argmin_{\Z}~ \Lagr{\T(k+1), \Z, \Y(k)} \text{ and } \label{eq:updateZ} \\
     \text{Dual update: } &        
        \Y(k + 1) = \argmin_{\Y}~ \Lagr{\T(k+1), \Z(k+1), \Y}, \nonumber 
\end{align}
where $k$ is the iteration step count. 
We find that the update step for $\Y(k+1)$ has the analytical solution
\begin{equation}
        \Y_i(k + 1) = \Y_i(k)  + \left[\T_i(k+1) - \Z_i(k+1) \right] \quad \text{for }i=1,2,\ldots,m. \label{eq:updateY} 
\end{equation}
The update steps for $\T(k+1)$ and $\Z(k+1)$ have to be solved numerically. The first has a known solution \citep{Danaher2014}. The latter is more involved and is discussed in Section \ref{sec:updateZ}.  

The algorithm iterates until convergence is reached. To guarantee convergence to an optimal solution, the constraint $\mb{\Theta}_i - \mb{Z}_i = \mathbf{0}$ must be satisfied while the augmented Lagrangian forms \eqref{eq:updateTheta} and \eqref{eq:updateZ} have to be minimized. In each iteration, the minimization constraint on the dual variable is satisfied, which ensures that the constraint update is satisfied by construction \citep{Boyd2010, Monti2014}. The likelihood update step holds asymptotically as $k \rightarrow \infty$ using the following stopping criterion \citep{Danaher2014, Wu2019}: 
\begin{equation}
    \sumim \norm{\T_i(k + 1) - \T_i(k)}_1 / \sumim \norm{\T_i(k)}_1 < \epsilon.
    \label{eq:stoppingcriterion}
\end{equation}
As a default, we choose $\epsilon = 10^{-5}$ as our convergence threshold and control the step size by setting $\rho = 1$.

\subsection{Update Step for $\T$} \label{sec:updateStepTheta}
The update step for the precision matrices $\T(k+1)$ in eq.~\eqref{eq:updateTheta} can be divided into $m$ distinct problems, one for each precision matrix: 
\begin{equation}
  \begin{split}
     \T_i(k+1) &= \argmin_{\T_i}~ \Lagr{\T_i, \Z_i(k), \Y_i(k)}   \\
      &= \argmin_{\T_i}~ \frac{n_i}{2}  \left[\text{trace}\left( \widehat{\Cov}_i \T_i \right) - \log \text{det}\left(\Cov_i \right) \right] + \\
      &\qquad  (\rho/2) \normF{\Y_i(k) + \left[\T_i - \Z_i(k) \right]}
  \end{split}
  \label{eq:updateThetaLagragian}
\end{equation}
for $i = 1, 2, \ldots, m$. The precision matrix $\T_i(k+1)$ can be determined numerically. 
Due to $\rho > 0$, one must choose $\T_i$ in such a way that $\T_i$ both minimizes the log-likelihood and is in the proximity of $\Z_i(k)$. The degree to which this must be enforced depends on both $\rho$ and $\Y_i(k)$. Others have shown that eq.~\eqref{eq:updateThetaLagragian} can be solved numerically by rescaling an eigenvalue decomposition (see \citet{Witten2009} and \citet{Monti2014}). More precisely, let $\b{Q}_i\mb{\Gamma}_i\mb{Q}_i^{\top}$ be the eigenvalue decomposition of $\widehat{\Cov}_i - (\rho/n_i)\left[\Z_i(k) - \Y_i(k) \right]$, where $\b{Q}_i$ the columns are the eigenvectors and $\b{\Gamma}_i = \text{diag}(\gamma_{1}, \gamma_{2}, \ldots, \gamma_{p})$ is a diagonal matrix with the corresponding eigenvalues on the diagonal. We rescale this matrix such that $\wt{\b{\Gamma}}_i = \text{diag}(\wt{\gamma}_{i1}, \wt{\gamma}_{i2}, \ldots, \wt{\gamma}_{ip})$ where $\wt{\gamma}_{ij} = \dfrac{n_i}{2 \rho} \left( (\gamma_{ij}^2 + 4 \rho / n_i)^{\tfrac{1}{2}}  - \gamma_{ij}\right)$ for $j = 1, 2, \ldots, p$. The update step from eq.~\eqref{eq:updateThetaLagragian} then equals 
\begin{equation}
    \T_{i}(k + 1) = \b{Q}_i\widetilde{\mb{\Gamma}}_i\mb{Q}_i^{\top}.
    \label{eq:solutionTheta}
\end{equation}
Note that each $\T_{i}(k+1)$ is symmetric. Furthermore, 
all $\wt{\gamma}_i$ are strictly positive, ensuring that each $\T_{i}(k+1)$ is positive definite.

The computationally most challenging part of the update step is the eigenvalue decomposition of a $(p \times p)$-matrix with a complexity of  $\mathcal{O}(p^3)$. Since this is needed for each of the $m$ graphs, the $\T$-update step's complexity is $\mathcal{O}(m p^3)$. 

\subsection{Update Step for $\Z$} \label{sec:updateZ}

The minimization of the augmented Lagragian with respect to $\Z$ in eq.~\eqref{eq:updateZ} is equal to 
\begin{equation}
    \begin{split}
        \Z(k+1) & = \argmin_{\Z} \Lagr{\T(k + 1), \Z, \Y(k+1)} \\ 
        & = \argmin_{\mb{Z}}~ (\rho/2)\sum_{i =1}^m \normF{\Z_i - \T_i(k+1) - \Y_{i}(k)} + \lambda_1 \sum_{i = 1}^m \normOff{\Z_i} \\ 
        & \qquad \qquad \qquad + \lambda_2 \sumij w_{ij} \normOff{\Z_i - \Z_j}.
    \end{split}
    \label{eq:zupdate}
\end{equation}
In contrast to $\T(k+1)$, for which we could divide the update step into $m$ distinct problems, we need to solve the problem for $\Z(k+1)$ jointly due to the smoothness term $\normOff{\Z_i - \Z_j}$. However, we can divide the problem into multiple, independent optimization problems, one for each node pair $\{s,t\}$, corresponding to a potential edge in the graphs. Let $\{s,t\}$ be the potential edge between nodes $X_s$ and $X_t$. Eq.~\eqref{eq:zupdate} can then be written as 
\begin{equation*}
    \begin{split}
        \Z(k+1) & = \argmin_{\Z}~ (\rho/2) \sumim \sum_{s, t} \left[z_{st}^{(i)} - \theta_{st}^{(i)}(k + 1) - y_{st}^{(i)} (k) \right]^2  \\ &  \qquad \qquad \qquad  + \lambda_1 \sumim \sum_{s \neq t} |z_{st}^{(i)}|  
        + \lambda_2 \sumij \sum_{s \neq t} w_{ij} |z^{(i)}_{st} - z^{(j)}_{st}|,
    \end{split}
    \label{eq:zupdateedge}
\end{equation*}
which is completely separable with respect to each node pairs $\{s,t\}$. Let 
\begin{align*}
    \bm{\theta}_{st} & = \left(\theta_{st}^{(1)}, \theta_{st}^{(2)}, \ldots, \theta_{st}^{(m)}\right)^\top, \\
    \bm{z}_{st} & = \left(z_{st}^{(1)}, z_{st}^{(2)}, \ldots, z_{st}^{(m)}\right)^\top, \quad \text{and} \\  
    \bm{y}_{st} & = \left(y_{st}^{(1)}, y_{st}^{(2)}, \ldots, y_{st}^{(m)}\right)^\top 
\end{align*}
be $m$-dimensional vectors with the $(s,t)$ entries of the matrices $\T$, $\Z$ and $\Y$, respectively. There is an analytical solution for the diagonal elements of $\bm{Z}(k+1)$, specifically:
\begin{equation*}
    \bm{z}_{ss}(k+1) = \argmin_{\bm{z}_{ss} \in \R^m}~ \norm{\bm{z}_{ss} - \bm{\theta}_{ss}(k+1) - \bm{y}_{ss}(k)}_2^2 = \bm{\theta}_{ss}(k+1) + \bm{y}_{ss}(k). 
\end{equation*}
For $s \neq t$, we can update each $(s,t)$ entry by solving
\begin{equation*}
    \begin{split}
        \bm{z}_{st}(k+1) & = \argmin_{\bm{z}_{st} \in \R^m}~ (\rho/2) \norm{\bm{z}_{st} - \bm{\theta}_{st}(k+1) - \bm{y}_{st}(k)}_2^2 \\ 
        & \qquad \qquad + \lambda_1 \norm{\bm{z}_{st}}_1 + \lambda_2 \sum_{i < j} w_{ij} |z_{st}^{(i)} - z_{st}^{(j)}|.  
    \end{split}
    \label{eq:updatezindividualedge}
\end{equation*}
We can express this problem as a \emph{weighted Fused LASSO Signal Approximator} (wFLSA). For ease of notation, let us define $\bm{\beta} = \bm{z}_{st}$ and ${\bm{y} = \bm{\theta}_{st}(k+1) + \bm{y}_{st}(k)}$; we then find that our problem can be written as 
\begin{equation}
    \bm{\beta}(k+1) = \argmin_{\bm{\beta} \in \R^m}~ \frac{1}{2} \normQuad{\bm{y} - \bm{\beta}} + \eta_1  \norm{\bm{\beta}}_1 + \eta_2 \sumij w_{ij} |\beta_i - \beta_j| 
    \label{eq:generalizedlasso}
\end{equation}
where $\eta_1 = \lambda_1 / \rho$ and $\eta_2 = \lambda_2 / \rho$. 
Since the wFLSA was not solved before, we developed an ADMM algorithm specifically to this end, see \citet{dijkstra2024spinoff}. The complexity of this algorithm is in the order of $\bigO(m^2)$.

\begin{sidewaysfigure}
    \centering
    \includegraphics[width=\textwidth]{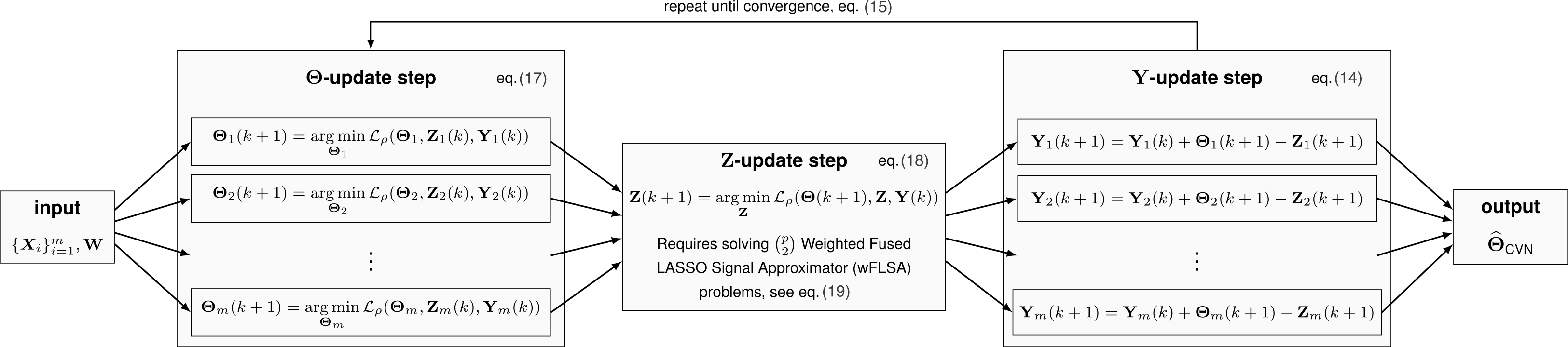}
    \caption{A graphical representation of the algorithm for estimating the CVN model, see eq.~\eqref{eq:cvn}. The input is the raw data set and the weight matrix; the output is the estimated CVN model. The ADMM consists of three updates steps for $\T$, $\Z$ and $\Y$ that are repeated till convergence. Update step $\T$ and $\Y$ can be subdivided into $m$ independent step. During the $\Z$-update step, $\binom{p}{2}$ wFLSA problems need to be solved, one for each potential edge.}
    \label{fig:algorithm}
\end{sidewaysfigure}

\subsection{Pseudo-algorithm for the CVN model} 

Combining the results presented in the previous section, leads us to the following procedure for estimating CVN models. A graphical representation can be found in Figure~\ref{fig:algorithm}. 

\begin{enumerate}
    \item Compute the empirical covariance matrices $\widetilde{\Cov}_i = ( \sigma_{st}^{(i)})_{p \times p} = n_i^{-1} \sum_{j = 1}^{n_i} \bm{x}_j^{(i)} \left(\bm{x}_j^{(i)} \right)^\top$ with $i = 1, 2, \ldots, m$. 
    \item{Initialize 
            $\T_i(1) = \text{diag}(\widetilde{\Cov}_i)^{-1}$ and $\b{Z}_i(1) = \b{Y}_i = \b{0}$. Alternatively, one can use a warmstart where $\T_i(1) = \argmin_{\T_i} \ell_i (\T_i) + \lambda_1 \normOff{\T_i}$ is the GLASSO estimate for the individual graphs $i = 1, 2, \ldots, m$.     
    }
    \item{Update $\T$. For each graph $i = 1,2,\ldots,m$, do:
        \begin{enumerate}
            \item Compute the eigenvalue decomposition $\b{Q}_i \b{\Gamma}_i \b{Q_i}^\top$ of $\widetilde{\Cov}_i - (\rho/n_i)\left[\Z_i(k) - \Y_i(k) \right]$; 
            \item Let $\widetilde{\b{\Gamma}}_i = \text{diag}(\widetilde{\gamma}_{i1}, \widetilde{\gamma}_{i2}, \ldots, \widetilde{\gamma}_{ip})$ where $\widetilde{\gamma}_{ij} = \frac{n_i}{2\rho} \left( \sqrt{\gamma_{ij}^2 + 4 \rho / n_i}  - \gamma_{ij} \right)$;
            \item Then $\T_{i}(k + 1) = \b{Q}_i\widetilde{\mb{\Gamma}}_i\mb{Q}_i^{\top}$.
        \end{enumerate}
    }
    \item{Update $\Z$ by going over all potential edges $\{s,t\}$:
        \begin{enumerate}
            \item{Let $\bm{\theta}_{st} = \left(\theta_{st}^{(1)}, \theta_{st}^{(2)}, \ldots, \theta_{st}^{(m)}\right)^\top$, 
    $\bm{z}_{st} = \left(z_{st}^{(1)}, z_{st}^{(2)}, \ldots, z_{st}^{(m)}\right)^\top$ and \\ $\bm{y}_{st} = \left(y_{st}^{(1)}, y_{st}^{(2)}, \ldots, y_{st}^{(m)}\right)^\top$; 
            }
            \item{Update the diagonal: $\bm{z}_{ss}(k+1) = \bm{\theta}_{ss}(k+1) + \bm{y}_{ss}(k)$ for $s = 1,2,\ldots,p$;}
            \item{Update the off-diagonal entries $\bm{z}_{st}(k + 1)$ and $\bm{z}_{ts}(k + 1)$ for all $s < t$. Let $\bm{\beta} = \bm{z}_{st}(k + 1)$ and $\bm{y} = \bm{\theta}_{st}(k + 1) + \bm{y}_{st}(k)$, then 
            $$
            \bm{z}_{st}(k+1) = \argmin_{\bm{\beta} \in \R^m}~ \frac{1}{2} \normQuad{\bm{y} - \bm{\beta}} + \eta_1  \norm{\bm{\beta}}_1 + \eta_2 \sumij w_{ij} |\beta_i - \beta_j|  
            $$
            is the corresponding wFLSA problem, where $\eta_1 = \lambda_1 / \rho$ and $\eta_2 = \lambda_2 / \rho$. Solve this using the algorithm presented in \citet{dijkstra2024spinoff}. 
            }
            
        \end{enumerate}
    }
    \item{Update $\Y$: $\Y_i(k + 1) = \T_i(k+1) - \Z_i(k+1) + \Y_i(k)$ for 
 $i=1,2,\ldots,m$.}
    \item{In case the stopping criterion   
        $
            \sumim \norm{\T_i(k + 1) - \T_i(k)}_1 / \sumim \norm{\T_i(k)}_1 < \epsilon
        $ has been met, return the estimate of the CVN: $\Tcvnest = \{\T_{i}(k+1) \}_{i = 1}^m$. Otherwise, repeat steps 3 to 5.
    }
\end{enumerate}
An implementation of this algorithm is publicly available as an \texttt{R}~package at\\  \texttt{www.github.com/bips-hb/CVN}.

\section{Algorithmic Complexity} \label{sec:computationalComplexity}

Here, we determine the computational complexity of the CVN algorithm presented in the previous section. We first consider each of the update steps separately and combine the results at the end. 

The $\T(k+1)$-update step can be decomposed into $m$ distinct steps, each associated with a specific graph $i = 1,2,\ldots,m$, as illustrated in Figure~\ref{fig:algorithm}. The most computationally demanding aspect of the $\T_i(k+1)$ update step, see eq.~\eqref{eq:updateTheta}, involves performing the eigendecomposition of a $(p \times p)$-dimensional matrix, a task known to be on the order of $\bigO(p^3)$. The entire update step requires, thus, in the order of $\bigO(mp^3)$ operations. 

In the case of the $\Y(k+1)$ update step, see eq.~\ref{eq:updateY}, the task entails summing three $(p \times p)$-dimensional matrices for each graph $i = 1,2,\ldots,m$. This operation, thus, requires $\bigO(m p^2)$ operations.

The $\Z(k+1)$-update step requires one to solve a wFLSA problem for each of the $\binom{p}{2}$ potential edges in the graph. We know that the complexity of solving a single wFLSA problem is $\bigO(r m^2)$, where $r$ is the (average) number of iterations needed for the algorithm to converge \citep{dijkstra2024spinoff}. The overall complexity of the $\Z(k+1)$-update step is, thus, $\bigO\left(\binom{p}{2} \cdot r m^2\right) = \bigO(rm^2p^2)$.

Combining these results, yields the computational complexity of the ADMM algorithm for solving CVN models as
$$
    \bigO\Big(R \cdot \big(\underbrace{mp^3}_{\T(k+1)\text{-update}} + \underbrace{mp^2}_{\Y(k+1)\text{-update}} + \underbrace{rm^2p^2}_{\Z(k+1)\text{-update}} \big)\Big) = \bigO\left(R r m^2p^3\right),
$$
where $R$ is the number of iterations needed for the ADMM to satisfy the stopping criterion in eq.~\eqref{eq:stoppingcriterion}, $r$ denotes the average iterations needed for the wFLSA algorithm to converge \citep{dijkstra2024spinoff}, $m$ is the number of graphs, and $p$ is the number of variables.

\section{Interpolation} \label{sec:interpolation}

In this section, we propose a method for interpolating a graph for which there are no observations based on an estimated CVN model with $m$ graphs. We denote the graph we want to interpolate by $G_{m + 1}$. We are concerned solely with the existence of edges and not values of the entries in the precision matrix $\T_{m+1}$. Approaches in the literature often rely on the so-called `AND' and `OR' rules (see, for example, \citet{Hallac2017}). In case of the former, the interpolated graph has the edge $\{s,t\}$  if and only if \emph{all} $m$ graphs in the CVN model have the edge $\{s,t\}$. In case of the `OR' rule, the interpolated graph has the edge $\{s,t\}$ if at least one of the other estimated graphs has that edge. Note that with this approach, it is difficult to control the level of sparsity and take different levels of similarity between the graphs into account. 

We interpolate graph $G_{m + 1}$ given the estimated CVN model $\Tcvnest$ by solving
\begin{equation*}
    \widehat{\T}_{m + 1} = \argmin_{\T_{m+1} \succ 0} \ell_{m+1}(\T_{m+1}) + \lambda_1 \normOff{\T_{m+1}} + \lambda_2 \sumim \omega_i \normOff{\widehat{\T}_i - \T_{m+1}} 
\end{equation*}
where $\bm{\omega} = (\omega_1, \omega_2, \ldots, \omega_m)^\top$ are the \emph{smoothing coefficients} which reflect the level of similarity between the individual graphs of the CVN and the interpolated graph. Note that, since there are no observations for the graph $G_{m+1}$, the corresponding log-likelihood is equal to $\ell_{m+1}(\T_{m+1}) = 0$, see eq.~\eqref{eq:likelihood}. Similar to the $\Z$-update step, we can divide the problem into $\binom{p}{2}$ separate optimization problems, one for each potential edge $\{s,t\}$. In fact, we can write the function as 
\begin{equation*}
    \begin{split}
        \widehat{\T}_{m + 1} & = \argmin_{\T_{m+1} \succ 0}\ \lambda_1 \sum_{s,t} |\theta_{st}^{(m+1)}| + \lambda_2 \sumim \omega_i \sum_{s,t} |\widehat{\theta}_{st}^{(i)} - \theta_{st}^{(m+1)}|  \\ 
        & = \argmin_{\T_{m+1} \succ 0}\ \sum_{s,t} \left[ \lambda_1 |\theta_{st}^{(m+1)}| + \lambda_2 \sumim w_{i} |\widehat{\theta}_{st}^{(i)} - \theta_{st}^{(m+1)}| \right].
    \end{split}
\end{equation*}
We, thus, need to solve for each potential edge $\{s,t\}$:
\begin{equation*}
    \wh{\theta}_{st}^{(m+1)} = \argmin_{\theta_{st}^{(m+1)} \in \mathds{R}} \lambda_1 |\theta_{st}^{(m+1)}| + \lambda_2 \normsimple{\bm{\omega}^\top \widehat{\bm{\theta}}_{st} - \theta_{st}^{(m+1)}}_1.
    \label{eq:interpolationSingleEdge}
\end{equation*}
This function is convex (see Appendix~\ref{sec:appendix:interpolation}) and can be solved using any derivative-free search method. In this context, we employ Brent's algorithm \citep{brent2013algorithms}. 
The sign of $\widehat{\theta}_{st}^{(m+1)}$ is inconsequential, as our primary objective is to establish the presence or absence of the edge $\{s,t\}$ in the interpolated graph, i.e., $\widehat{\theta}_{st}^{(m+1)} \neq 0$. 
Note that the smoothing coefficients $\omega_1, \omega_2, \ldots, \omega_m$ need to be chosen a priori, which requires a certain level of prior knowledge.  

\section{Tuning Parameter Selection} \label{sec:tuningParameterSelection}

After fitting the CVN model, the next step involves selecting suitable values for the tuning parameters: $\lambda_1$, which governs the sparsity of the graphs, and $\lambda_2$, responsible for regulating the smoothness or similarity between the graphs. Since having prior knowledge about appropriate values for the tuning parameters is rare, one often opts for a data-driven approach. Numerous methods exist for tuning parameter selection, and in this context, we explore two frequently used measures: the Akaike Information Criterion (AIC) and Bayesian Information Criterion (BIC). However, it is important to note that in our scenario, these criteria can only be approximated, see \citet{Danaher2014}.

Let $\widehat{\T}_i(\lambda_1, \lambda_2)$ denote the estimator of the precision matrix for the $i$-th graph for specific values of $\lambda_1$ and $\lambda_2$. The AIC can be approximated as
\begin{equation}
    \text{AIC}(\lambda_1, \lambda_2) = \sum_{i = 1}^m \ell_i (\widehat{\T}_i(\lambda_1, \lambda_2)) + 2 \normsimple{\widehat{\T}_i(\lambda_1, \lambda_2)}_0, 
    \label{eq:definitionAIC}
\end{equation}
where $\norm{\cdot}_0$ denotes the $L_0$-norm, equivalent to the number of non-zero entries in the matrix \citep{Danaher2014}. On the other hand, the BIC, taking into account the number of observations for the graphs, tends to result in sparser models than the AIC:
\begin{equation}
     \text{BIC}(\lambda_1, \lambda_2) = \sum_{i = 1}^m \ell_i (\widehat{\T}_i(\lambda_1, \lambda_2)) + 2 \log(n_i) \normsimple{\widehat{\T}_i(\lambda_1, \lambda_2)}_0, 
     \label{eq:definitionBIC}
\end{equation}
where $n_i$ is the number of observations for the $i$-th graph. 
The optimal values for $\lambda_1$ and $\lambda_2$ are those that minimize either the approximated AIC or BIC. 
In our simulation study, see Section~\ref{sec:simulation}, we consider both information criteria. Note that while AIC and BIC are commonly used, they are not the sole metrics for selecting the optimal tuning parameters based on the data. Other methods are cross-validation, bootstrap \citep{hastie2009elements}, and alternative information criteria like the extended BIC \citep{chen2008extended} and Predictive Information Criteria \citep{ando2007bayesian}. Here, we opt for AIC and BIC due to the computational efficiency, which is advantageous for a computationally intensive algorithm such as the ADMM propsed here.

\section{An Alternative Tuning Parameterization} \label{sec:alternativeTuningParameterization}  

Due to the computational complexity of the algorithm  exploring an extensive range of values for the tuning parameters $\lambda_1$ and $\lambda_2$ is impractical, and, as mentioned above, obtaining expert or prior knowledge about `optimal' values  for these parameters is challenging. 

We introduce an alternative parameterization for the CVN model. Instead of using $\lambda_1$ and $\lambda_2$, we propose employing two different parameters, $\gamma_1$ and $\gamma_2$, which we define in this section. The benefit of this new parameterization is that the interpretation of these values do \emph{not} heavily depend on the number of graphs ($m$) and/or the number of variables ($p$).

Suppose a model has been fitted and the best values for the tuning parameters $\lambda_1$ and $\lambda_2$ have already been selected based on some criterion, see the previous section. Now, consider a scenario where the same data set is used, but a number of variables are added or removed (thereby changing $p$), and/or other external covariates are added or removed (changing the number of graphs $m$). Due to the way the model is defined, see eq.~\eqref{eq:cvn}, the interpretation of the penalty terms $\lambda_1$ and $\lambda_2$ change, i.e., the optimal values for the first model provide little information about the best values for the tuning parameters for the new model. 

Note that this challenge is not unique to the CVN model; it extends to other penalized regression problems as well. However, this challenge is typically overlooked for two main reasons. Firstly, it is in other situations uncommon for the considered variables to change. Secondly, the associated algorithm is usually computationally light, making it straightforward to reapply to new datasets.
Nevertheless, we would like to point out that the alternative tuning parameterization proposed here may be useful for other LASSO-related methods as well. 

In the existing parameterization, penalties are imposed on the sum of the $L_1$-norm of the precision matrices, governed by $\lambda_1$, and the sum of the $L_1$-norm of the differences between the precision matrices, regulated by $\lambda_2$, see eq.~\eqref{eq:cvn}. We propose a modification where we penalize each individual edge in the CVN model to regulate sparsity, and each pair of edges in the CVN model to control the level of smoothness. We denote the new parameters as $\gamma_1$, regulating sparsity, and $\gamma_2$, managing smoothness.

Recall that the total number of potential edges in the CVN model is $m p (p - 1)$, and the total number of pairs of potential edges is $m(m-1)p(p-1)$. We define $\gamma_1$ and $\gamma_2$ as follows:
\begin{equation*}
    \gamma_1 = 2\frac{\lambda_1}{mp(p-1)} \qquad \text{and} \qquad \gamma_2 = 4 \frac{\lambda_2}{m(m-1)p(p-1)}.
\end{equation*}
Consequently, the sparsity penalty term can be expressed as
\begin{equation*}
    \lambda_1 \sumim \normOff{\T_i} = \sumim \sum_{s < t} \gamma_1 |\theta_{st}^{(i)}|.
\end{equation*}
In other words, $\gamma_1$ regulates how much an individual edge in a graph is penalized. Consequently, alterations in the number of edges have a relatively modest impact on this tuning parameter in comparison to $\lambda_1$.
Similarly, the smoothness penalty term of the CVN model can be written in terms of $\gamma_2$:
\begin{equation*}
    \lambda_2 \sum_{i < j} \normOff{\T_{i} - \T_{j}} = \sum_{i < j} w_{ij} \sum_{s,t} \gamma_2 |\theta_{st}^{(i)} - \theta_{st}^{(j)}|.
\end{equation*}
Here, $\gamma_2$ is used to penalize a single pair of edges. Once more, changing the number of graphs and edges has little effect on the interpretation of this tuning parameter. 

The idea that optimal values for $\gamma_1$ and $\gamma_2$ are consistent across various datasets hinges on the assumption that both the sparsity of the graphs and their smoothness will remain relatively stable across those datasets.


\section{Simulation}\label{sec:simulation}

In this simulation study, we consider a CVN model with two external covariates, $U_1$ and $U_2$, each of which takes values from the set $\{1,2,3\}$ featuring nine graphs. The graph corresponding to $U_1 = i$ and $U_2 = j$ is denoted as $G_{ij} = G(U_1 = i, U_2 = j)$, and its associated adjacency, precision, and covariance matrices are similarly indexed as $\mathbf{A}_{ij}$, $\T_{ij}$ and $\mathbf{\Sigma}_{ij}$, respectively. Our simulation setup allows for varying the number of observations ($n$), the number of variables ($p$) and the the general graph structure, specifically Erd\"os-R\'enyi and Barabasi-Albert graphs \citep{newman2018networks} in our case. 
Additionally, we can control the extent to which the edges in the graph change with $U_1$ and $U_2$. Adjusting $n$ and $p$ is straightforward, as we will see later.
Our focus is first on how to employ a specific general graph structure and how to regulate the extent with which the graphs change with changes in the two external covariates.

\begin{figure}[h!]
    \centering
    \includegraphics[width=.5\textwidth]{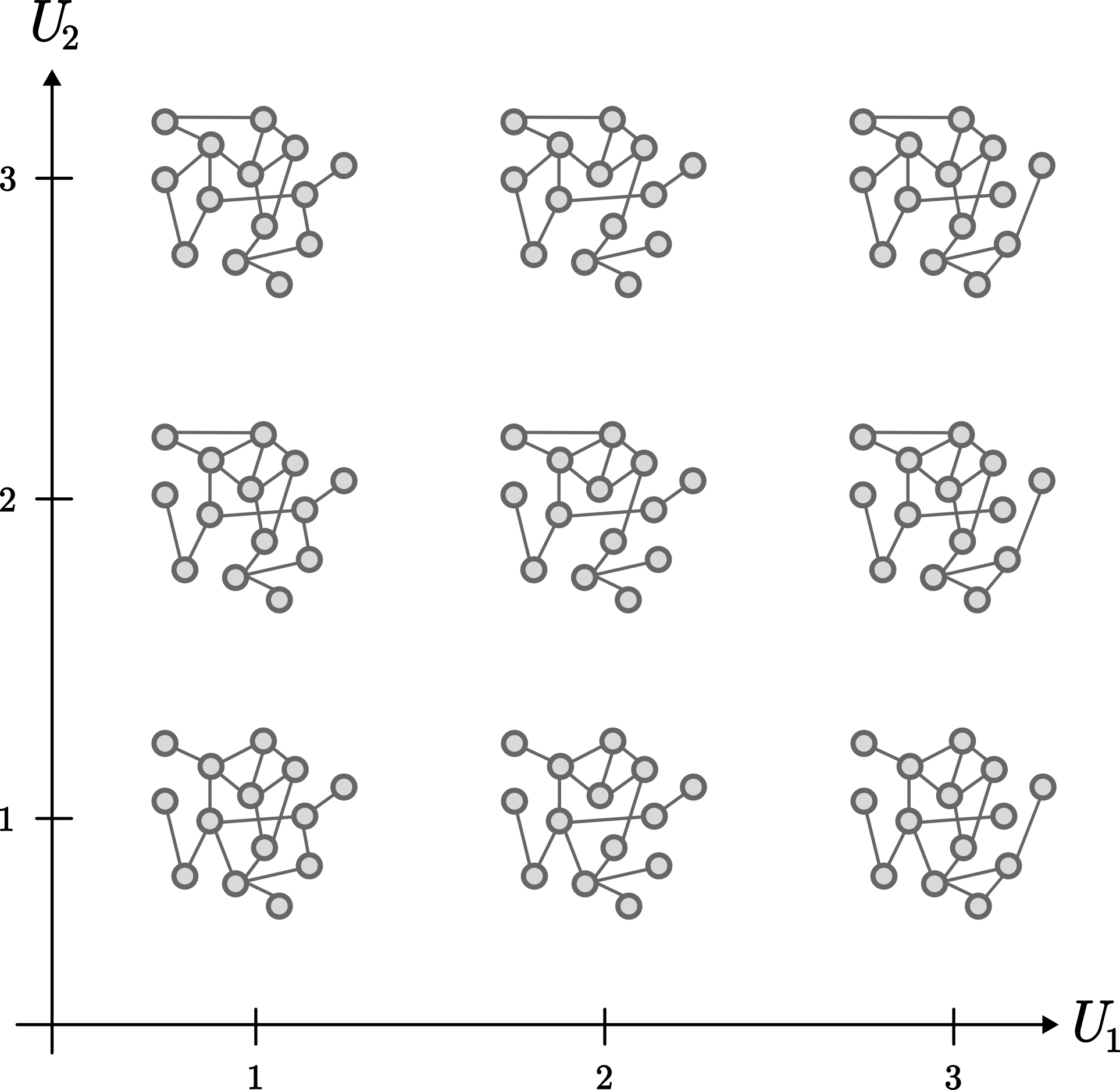}
    \caption{Illustration of the CNV model with two external covariates, $U_1$ and $U_2$, as used in the simulation study. The values of $U_1$ and $U_2$ are shown on the horizontal and vertical axes, respectively. The `starting graph' $G_{11}$ is located in the lower-left corner, followed by graph $G_{21}$ to its right, and so on, ending with graph $G_{33}$ is in the upper-right corner.}
    \label{fig:3x3gridOfGraphsSimulation}
\end{figure}

We can represent our case as a grid of graphs, as illustrated in Figure~\ref{fig:3x3gridOfGraphsSimulation}, with $U_1$ and $U_2$ positioned along the horizontal and vertical axes, respectively. Graph $G_{11}$ is located in the bottom-left corner, followed by graph $G_{21}$ to its right, and so forth, concluding with graph $G_{33}$ in the upper-right corner.  We designate graph $G_{11}$ as the \emph{starting graph}; it is randomly generated according to a specific graph model. In this context, we consider two types:
\begin{enumerate}
    \item The Erd\"os-R\'enyi (ER) graph, which has a single parameter $\pi$, representing the probability of the existence of any given edge. The expected density of the graph, defined as the number of edges divided by the total number of possible edges (i.e., $p(p-1)/2$), is expressed as \citep{newman2018networks}
\begin{equation*}
    \mathds{E}_{\text{ER}} \left[ |E| \right] = \pi.
    \label{eq:expectedDensityErdosRenyiGraph}
\end{equation*}
Here, $|E|$ denotes the cardinality of the edge set, and $\mathds{E}_{\text{ER}}$ signifies the expectation under the ER model. The relevance of the expected density of the graph becomes apparent later;
    \item The Barabasi-Albert (BA) graph,  which is a scale-free graph. The degrees of the graph follow a
power law distribution with an exponent of 3 \citep{newman2018networks}. The expected density of this graph type is approximately given by \citep{posfai2016network}
\begin{equation}
    \mathds{E}_{\text{BA}} \left[ |E| \right] \approx \frac{p - 1}{2},
    \label{eq:expectedDensityBarabasiAlbertGraph}
\end{equation}
where $\mathds{E}_{\text{BA}}$ represents the expectation under the BA model.
\end{enumerate}

\begin{figure}
\centering
\begin{subfigure}{.45\textwidth}
    \centering
    \includegraphics[width=\textwidth]{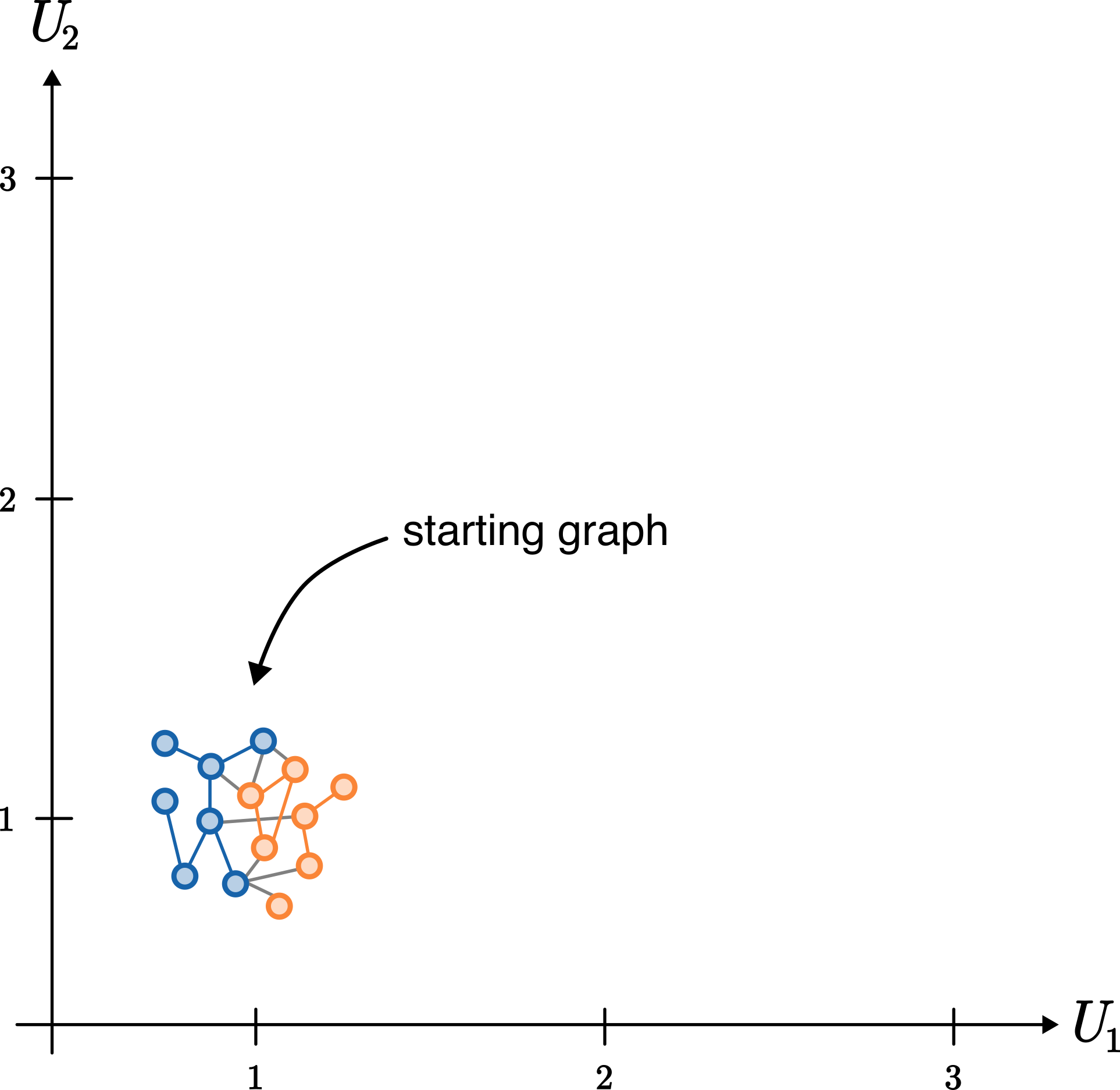}
    \caption{Dividing the starting graph $G_{11}$}\label{fig:simulationStartingGraph}
\end{subfigure}
    \hfill
\begin{subfigure}{.45\textwidth}
    \centering
    \includegraphics[width=\textwidth]{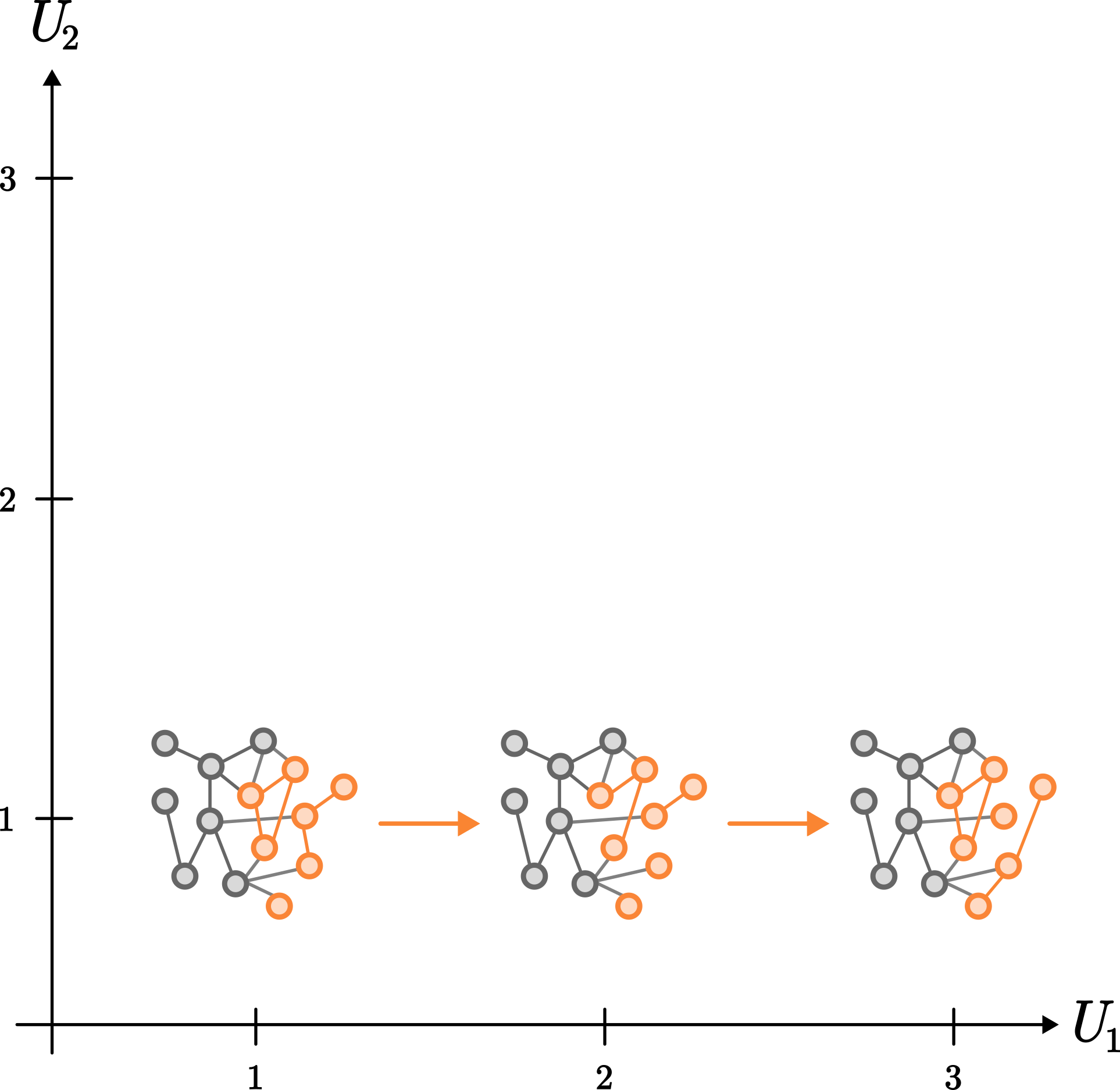}
    \caption{Changing the first subgraph}\label{fig:simulationChangesInU1Direction}
\end{subfigure}

\bigskip

\begin{subfigure}{.45\textwidth}
    \centering
    \includegraphics[width=\textwidth]{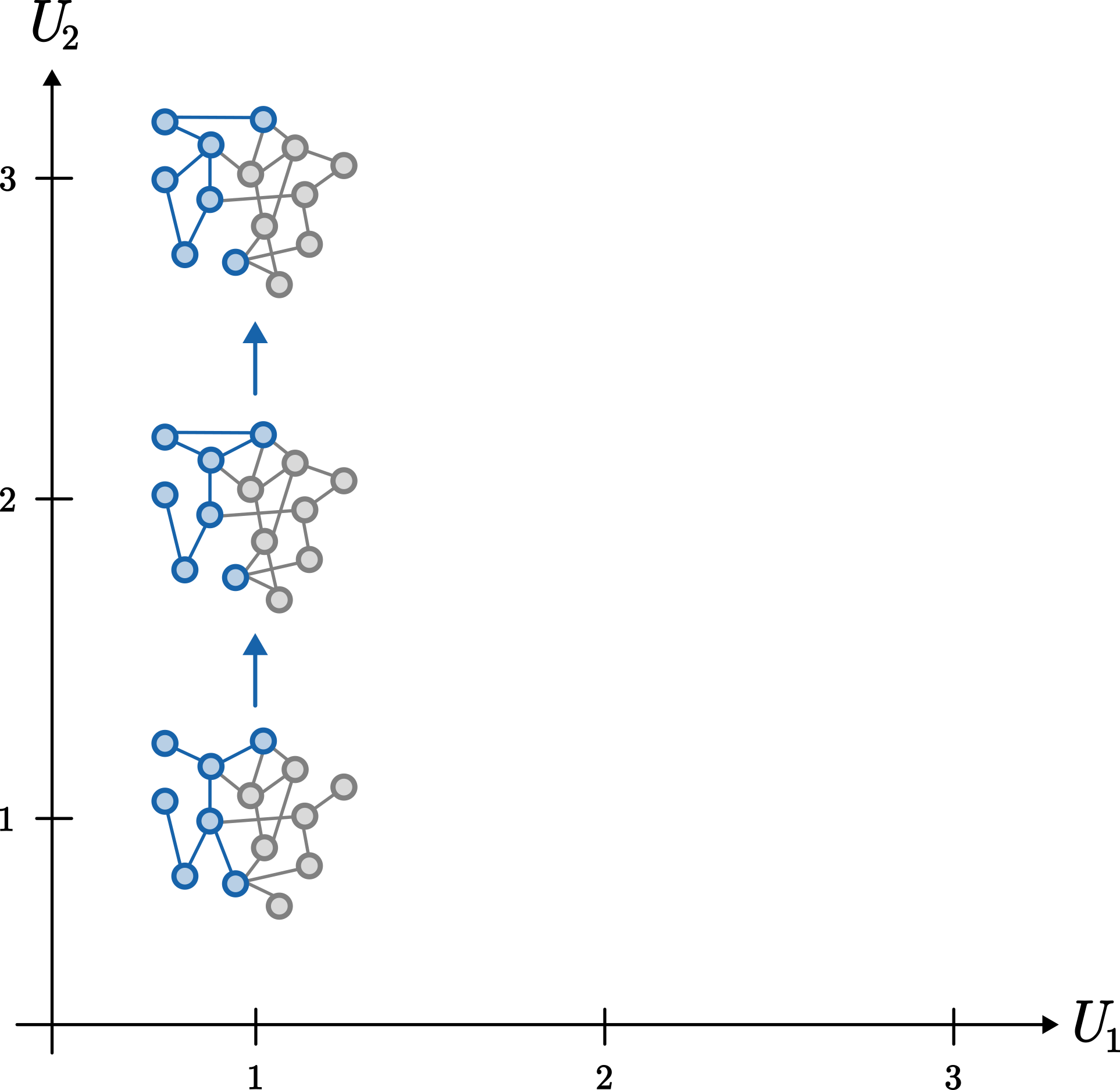}
    \caption{Changing the second subgraph}\label{fig:simulationChangesinU2Direction}
\end{subfigure}
    \hfill
\begin{subfigure}{.45\textwidth}
    \centering
    \includegraphics[width=\textwidth]{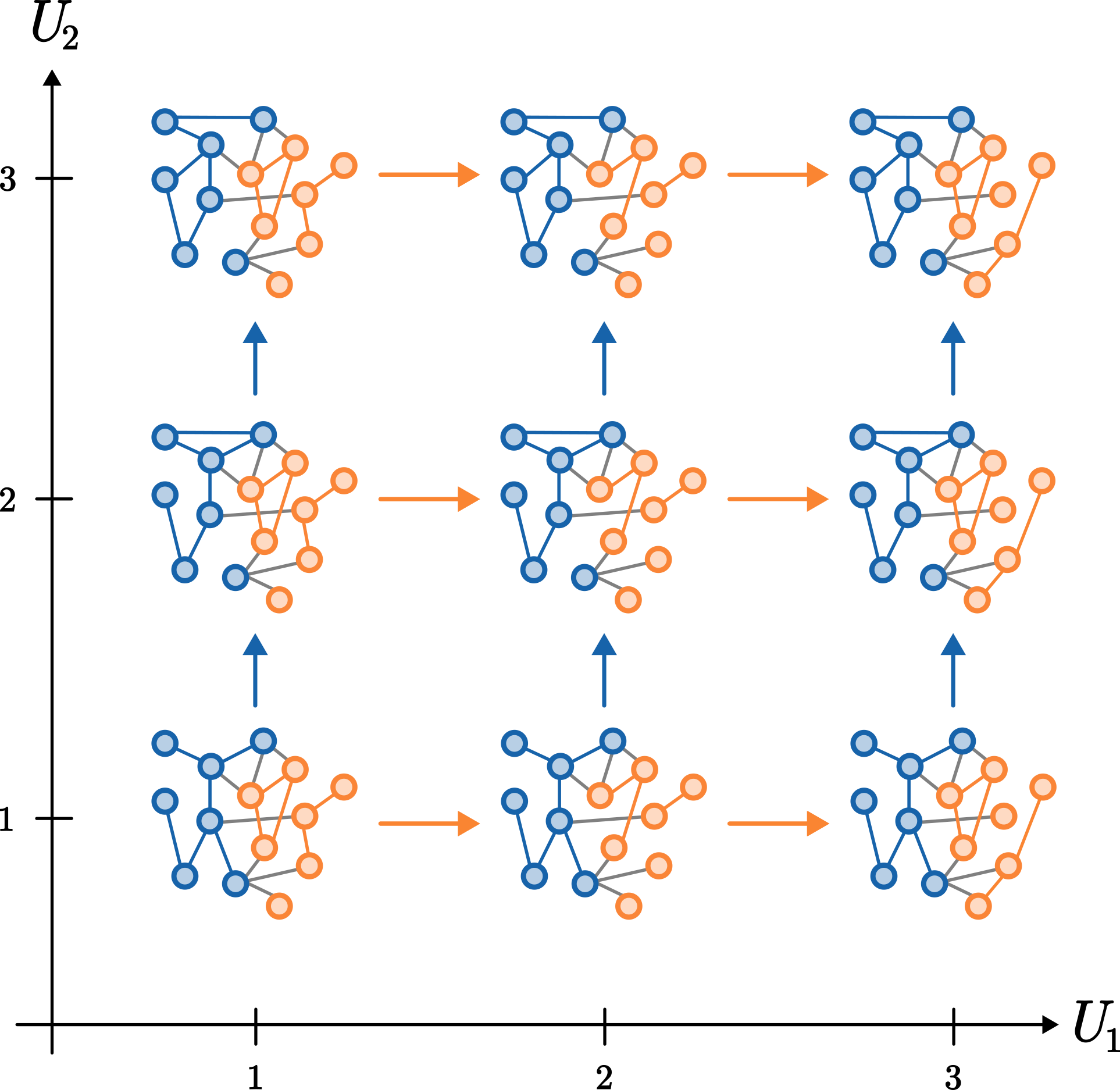}
    \caption{Combining the results}\label{fig:simulationCombiningChangingU1andU2Directions}
\end{subfigure}

\RawCaption{\caption{
An example illustrating the generation of the nine graphs used in the simulation study. (a) The starting graph is divided into two subgraphs (orange and blue). (b) The edges in the orange subgraph change with respect to $U_1$. (c) The edges in the blue subgraph change with respect to $U_2$. (d) Combining both modifications in the orange and blue subgraphs yields the nine graphs. Note that the edges between the two subgraphs (depicted in gray) remain unaltered.}
\label{fig:simulationCreatingTheGraphs}}
\end{figure}

As previously mentioned, we aim to control the extent to which graphs change with $U_1$ and $U_2$. To achieve this, we partition the starting graph into two equal-sized subgraphs. See, for an example, Figure~\ref{fig:simulationStartingGraph}, where the first subgraph is shown in orange and the second in blue. In cases where the number of variables $p$ is odd, the first and second subgraphs comprise $\lceil{p/2}\rceil$ and $\lfloor{p/2}\rfloor$ nodes, respectively, where $\lceil\cdot\rceil$ and $\lfloor\cdot\rfloor$ denote the ceiling and floor functions. The adjacency matrix of the starting graph can be expressed in terms of these subgraphs as follows:
\begin{equation}
    \b{A}_{11} = \begin{bmatrix}
        \b{A}_1^1 & \b{A} \\ 
        \b{A}^\top & \b{A}_1^2
    \end{bmatrix}.
    \label{eq:shapeAdjacencyMatrixStartingGraphSimulationStudy}
\end{equation}
Here, $\mathbf{A}_1^1$ represents the adjacency matrix for the first subgraph, and $\mathbf{A}_1^2$ is the matrix for the second subgraph. Similarly, we divide the remaining eight graphs into subgraphs using the same structure:
\begin{equation*}
    \b{A}_{ij} = \begin{bmatrix}
        \b{A}_i^1 & \b{A} \\ 
        \b{A}^\top & \b{A}_j^2
    \end{bmatrix}
    \label{eq:generalShapeAdjacencyMatrixSimulationStudy}
\end{equation*}
for $i = 1,2,3$ and $j = 1,2,3$. 

When varying $U_1$, only the subgraph corresponding to $\mathbf{A}_i^1$ undergoes changes, while the remainder of the graph remains unaltered. Similarly, when varying $U_2$, only the matrix $\mathbf{A}_j^2$ undergoes changes. It is important to note that the edges between the two subgraphs, represented here as $\mathbf{A}$, remain constant throughout and form the \emph{core} of the network. 

After sampling the starting graph $G_{11}$ and obtaining $\mathbf{A}_1^1$ and $\mathbf{A}_1^2$, we derive $\mathbf{A}_2^1$ by randomly adding and removing $b_1$ edges. If there are no $b_1$ edges in the subgraph, all existing edges change. Subsequently, we obtain $\mathbf{A}_3^1$ by altering $b_1$ edges in $\mathbf{A}_2^1$.
A similar process is followed for the second subgraph: we obtain $\mathbf{A}_2^2$ by randomly adding and removing $b_2$ edges in the matrix $\mathbf{A}_1^2$. The matrix $\mathbf{A}_3^2$ is then created by modifying $b_2$ edges in the matrix $\mathbf{A}_2^2$.

We visually depict this process in Figure~\ref{fig:simulationCreatingTheGraphs}. Figure~\ref{fig:simulationStartingGraph} displays the starting graph $G_{11}$ with the first subgraph depicted in orange and the second in blue. Changes occur exclusively in the edges of the orange subgraph as $U_1$ varies, as illustrated in Figure~\ref{fig:simulationChangesInU1Direction}. Similarly, only edges in the blue subgraph undergo changes with variations in $U_2$, as shown in Figure~\ref{fig:simulationChangesinU2Direction}. Combining the changes made in the orange and blue subgraphs results in the nine graphs depicted in Figure~\ref{fig:simulationCombiningChangingU1andU2Directions}.

Focusing on the number of edges $b_1$ and $b_2$ that change with varying either $U_1$ or $U_2$ may be challenging to interpret, particularly for different values of $p$ or the number of edges. Therefore, we find it more meaningful to think in terms of the percentage of edges in the graph that change with $U_1$ and $U_2$. These percentages are denoted as $\pi_1$ and $\pi_2$. The number of edges changing with $U_1$ and $U_2$ are then given by
\begin{equation}
    b_1 = \text{round}\left( \pi_1 \mathds{E}[|E|] \right) \qquad \text{and} \qquad b_2 = \text{round}\left( \pi_2 \mathds{E}[|E|] \right),
    \label{eq:determineNumberEdgesChangeGivenPercentages}
\end{equation}
where $\text{round}(\cdot)$ rounds to the nearest integer. In other words, $\pi_1$ and $\pi_2$ represent the percentage of edges the graph is expected to have ($\mathds{E}[|E|]$) that change with $U_1$ and $U_2$, respectively. 

We translate the obtained adjacency matrices into nine precision matrices. Subsequently, these precision matrices are used to generate data using a normal distribution.
We adopt a methodology similar to the one outlined in \citep{Liu2017}. We define $\wt{\T}_{ij} = v \b{A}_{ij}$ as the adjacency matrix for $U_1 = i$ and $U_2 = j$, multiplied by the scalar $v > 0$. Let $\kappa_{ij}$ represent the smallest eigenvalue of the matrix $\wt{\T}_{ij}$. To ensure the precision matrix is positive definite, we `increase' the diagonal, specifically,
\begin{equation}
\T_{ij} = \wt{\T}_{ij} + \text{diag}\left( |\kappa_{ij}| + .1 + u \right),
\label{eq:precisionMatrixSimulationStudy}
\end{equation}
where $u > 0$ is a scalar as well. Throughout our simulation, we maintain $v = .4$ and $u = .1$, consistent with the values employed by \citet{Liu2017}. The corresponding covariance matrix is simply the inverse of the precision matrix, i.e., $\b{\Sigma}_{ij} = \T_{ij}^{-1}$, and is determined numerically. The data for each graph, $\b{X}_{ij}$, consists of $n$ independent samples from the $p$-dimensional normal distribution $\mathcal{N}(\b{0}, \b{\Sigma}_{ij})$.

In conclusion, our simulator follows the steps: 
\begin{enumerate}
    \item Specify the number of observations ($n$), parameters ($p$), initial graph type (ER or BA model), and the percentage of edges in the graphs changing in the $U_1$ and $U_2$ directions ($\pi_1$ and $\pi_2$); 
    \item Determine the number of changing edges $b_1$ and $b_2$ based on the percentages $\pi_1$ and $\pi_2$ using equation~\eqref{eq:determineNumberEdgesChangeGivenPercentages}; 
    \item Generate the starting graph $G_{11}$ with $p$ vertices following the selected model from step~1.
    \item Divide the starting graph $G_{11}$ into two subgraphs, see eq.~\eqref{eq:shapeAdjacencyMatrixStartingGraphSimulationStudy} and Figure~\ref{fig:simulationStartingGraph}; 
    \item Generate the adjacency matrix $A_2^1$ by randomly adding and removing $b_1$ edges from $A_1^1$, see Figure~\ref{fig:simulationChangesInU1Direction};
\item Generate $A_3^1$ by adding and removing $b_1$ edges from $A_2^1$ obtained in the previous step; 
\item Repeat the last two steps for the adjacency matrices $A_2^2$ and $A_3^2$ by changing $b_2$ edges, see Figure~\ref{fig:simulationChangesinU2Direction}; 
\item Steps 3--7 result in nine adjacency matrices, see Figure~\ref{fig:simulationCombiningChangingU1andU2Directions}. Determine the corresponding nine precision matrices using eq.~\eqref{eq:precisionMatrixSimulationStudy};
\item Numerically determine the covariance matrices (inverse of the precision matrices) from the previous step, and
\item Generate nine data sets ${\b{X}_{ij}}$, where $\b{X}_{ij}$ comprises $n$ independent draws from the $p$-dimensional normal distribution $\mathcal{N}(\b{0}, \b{\Sigma}_{ij}$).
\end{enumerate}
The simulator is publicly available as an \texttt{R} package at \url{www.github.com/bips-hb/CVNSim}.

\section{Simulation Set-up} \label{sec:simulationSetUp}

In our simulation study, we explore various parameter settings outlined in Table~\ref{tab:parameterSettingsSimulationStudy}. We set the number of variables, $p$, to 100 and vary the number of observations, $n$, from 100 to 200. Both scenarios are high-dimensional, as the number of edges in each graph is $\binom{p}{2} = 4{,}950$. As previously mentioned, we consider both the ER and BA models. For the ER model, the density $\pi$ is set to $10\%$. The sparsity of the BA model is constant, as given in eq.~\eqref{eq:expectedDensityBarabasiAlbertGraph}. We introduce variability in the degree to which graphs change in the $U_1$ and $U_2$ directions by setting $(\pi_1, \pi_2)$ to $(.1, .1)$, $(.1, .2)$, and $(.2, .2)$.

\begin{table}[h!]
    \centering
    \caption{Parameter settings for the simulation study}
\label{tab:parameterSettingsSimulationStudy}
    \begin{tabular}{l c c}
    \toprule
    \textbf{Description} & \textbf{Parameter} & \textbf{Value} \\ \midrule
     Number of variables & $p$ & 100 \\ 
    Number of observations & $n$ & 100, 200\\
    Density ER graph & $\pi$ & .1 \\ 
    Percentage edges changing with $U_1$ & $\pi_1$ & .1, .2 \\ 
    Percentage edges changing with $U_2$ & $\pi_2$ & .1, .2 \\ \bottomrule
    \end{tabular}
\end{table}

Using the CVN algorithm necessitates the specification of $\lambda_1$ and $\lambda_2$, or equivalently, $\gamma_1$ and $\gamma_2$ (see Section~\ref{sec:alternativeTuningParameterization}), along with the weight matrix $\b{W}$. In our simulation, we opt for $(\gamma_1, \gamma_2) \in \Gamma \times \Gamma$, where $\Gamma = \{10^{-5}, 5\cdot10^{-5}, 10^{-4}, 5\cdot10^{-4}, 10^{-3}, 5\cdot10^{-3}\}$. We explore three distinct weight matrices: 
\begin{enumerate}
    \item $\b{W}_0 = \bm{0}_{m \times m}$, the zero matrix. This corresponds to applying the GLASSO to each of the nine graphs individually, i.e., there is no smoothing; 
    \item $\b{W}_{\text{grid}}$, defined as
    \begin{equation}
        \b{W}_\text{grid} = \begin{bmatrix}
0 & 1 & 0 & 1 & 0 & 0 & 0 & 0 & 0 \\ 
1 & 0 & 1 & 0 & 1 & 0 & 0 & 0 & 0 \\ 
0 & 1 & 0 & 0 & 0 & 1 & 0 & 0 & 0 \\ 
1 & 0 & 0 & 0 & 1 & 0 & 1 & 0 & 0 \\ 
0 & 1 & 0 & 1 & 0 & 1 & 0 & 1 & 0 \\ 
0 & 0 & 1 & 0 & 1 & 0 & 0 & 0 & 1 \\ 
0 & 0 & 0 & 1 & 0 & 0 & 0 & 1 & 0 \\ 
0 & 0 & 0 & 0 & 1 & 0 & 1 & 0 & 1 \\ 
0 & 0 & 0 & 0 & 0 & 1 & 0 & 1 & 0 \\ 
\end{bmatrix}
\label{eq:weightMatrixGrid}
    \end{equation}
    representing the scenario where each graph is smoothed with its `adjacent' graphs, and
    \item $\b{W}_{\text{full}} = \b{1}_{m \times m} - \text{diag}(\b{1})$, a matrix of ones with zeros on its diagonal. In this case, each graph is equally smoothed with every other graph.
\end{enumerate}
See Figure~\ref{fig:simulationWeightMatrices} for a visual representation of the meta-graphs implied by these three weight matrices. 
Given the simulation process for the nine graphs, the second weight matrix, $\b{W}_\text{grid}$, mirrors the true underlying structure closest. We repeat the simulation for each simulation setting and weight matrix 20 times. 

As performance measure, we consider the $F_1$-score which we define here. Recall from the previous section that the true adjacency matrix is denoted as $\b{A}_{ij} = \left( a_{st}^{(ij)} \right)_{p \times p}$, and let $\widehat{\b{A}}_{ij}(\gamma_1, \gamma_2) = \left( \widehat{a}_{st}^{(ij)}(\gamma_1, \gamma_2)\right)_{p \times p}$ be the estimated adjacency matrix for $U_1 = i$, $U_2 = j$ and tuning parameters $\gamma_1$ and $\gamma_2$. 
The number of true positives (TP), false positives (FP) and false negatives (FN) are
\begin{equation*}
    \begin{split}
    \text{TP}(\gamma_1, \gamma_2) & = \sum_{i = 1}^3 \sum_{j = 1}^3 \left( \sum_{s < t} a_{st}^{(ij)}\cdot \widehat{a}_{st}^{(ij)}(\gamma_1, \gamma_2) \right),  \\   
    \text{FP}(\gamma_1, \gamma_2) & = \sum_{i = 1}^3 \sum_{j = 1}^3 \left( \sum_{s < t} \left(1 - a_{st}^{(ij)}\right) \cdot \widehat{a}_{st}^{(ij)}(\gamma_1, \gamma_2) \right) \text{ and }\\ 
    \text{FN}(\gamma_1, \gamma_2) & = \sum_{i = 1}^3 \sum_{j = 1}^3 \left( \sum_{s < t} a_{st}^{(ij)} \cdot \left(1 - \widehat{a}_{st}^{(ij)}(\gamma_1, \gamma_2) \right) \right). 
    \end{split}
\end{equation*}
Note that one needs to sum over all nine graphs. 
Using these definitions, we can express precision as follows:
\begin{equation*}
    \text{Precision}\left(\gamma_1, \gamma_2\right) = \frac{\text{TP}\left(\gamma_1, \gamma_2\right)}{\text{TP}\left(\gamma_1, \gamma_2\right) + \text{FP}\left(\gamma_1, \gamma_2\right)}.
\end{equation*}
And recall as: 
\begin{equation*}
    \text{Recall}\left(\gamma_1, \gamma_2\right) = \frac{\text{TP}\left(\gamma_1, \gamma_2\right)}{\text{TP}\left(\gamma_1, \gamma_2\right) + \text{FN}\left(\gamma_1, \gamma_2\right)}. 
\end{equation*}
The $F_1$-score is then simply the harmonic mean of the precision and recall: 
\begin{equation*}
    F_1\left(\gamma_1, \gamma_2\right) = 2 \cdot \frac{\text{Precision}\left(\gamma_1, \gamma_2\right) \cdot \text{Recall}\left(\gamma_1, \gamma_2\right)}{\text{Precision}\left(\gamma_1, \gamma_2\right) + \text{Recall}\left(\gamma_1, \gamma_2\right)}.
    \label{eq:definitionF1score}
\end{equation*}
The $F_1$-score requires selecting the tuning parameters $\gamma_1$ and $\gamma_2$ (or equivalently, $\lambda_1$ and $\lambda_2$). We consider all values of $(\gamma_1, \gamma_2) \in \Gamma \times \Gamma$. If the selection is based on the AIC or the BIC, see eq.~\eqref{eq:definitionAIC} and \eqref{eq:definitionBIC}, respectively, the $F_1$-scores are given by
\begin{equation}
    F_1^{\text{AIC}} = \min_{\left(\gamma_1, \gamma_2\right) \in \Gamma \times \Gamma} \text{AIC}\left(\gamma_1, \gamma_2\right) \quad \text{and} \quad  F_1^{\text{BIC}} = \min_{\left(\gamma_1, \gamma_2\right) \in \Gamma \times \Gamma} \text{BIC}\left(\gamma_1, \gamma_2\right).
    \label{eq:definitionF1scoreUsingAICorBIC}
\end{equation}
i.e., the scores obtained for the $\gamma_1$ and $\gamma_2$ values that minimize the AIC or the BIC. 
Additionally, we consider the oracle $F_1$-score, which represents the highest $F_1$-score achievable using the CVN algorithm if the true graphs are known. This score is given by
\begin{equation}
    F_1^\text{oracle} = \max_{\left(\gamma_1, \gamma_2\right) \in \Gamma \times \Gamma} F_1\left(\gamma_1, \gamma_2\right).
    \label{eq:F1Oracle}
\end{equation}
The oracle score serves as a benchmark to determine the best possible performance achievable when selecting the optimal tuning parameters.

\begin{figure}[h!]
\label{fig:simulationWeightMatrices}
    \centering
    \begin{subfigure}{.3\textwidth}
    \centering
    \includegraphics[width=\textwidth]{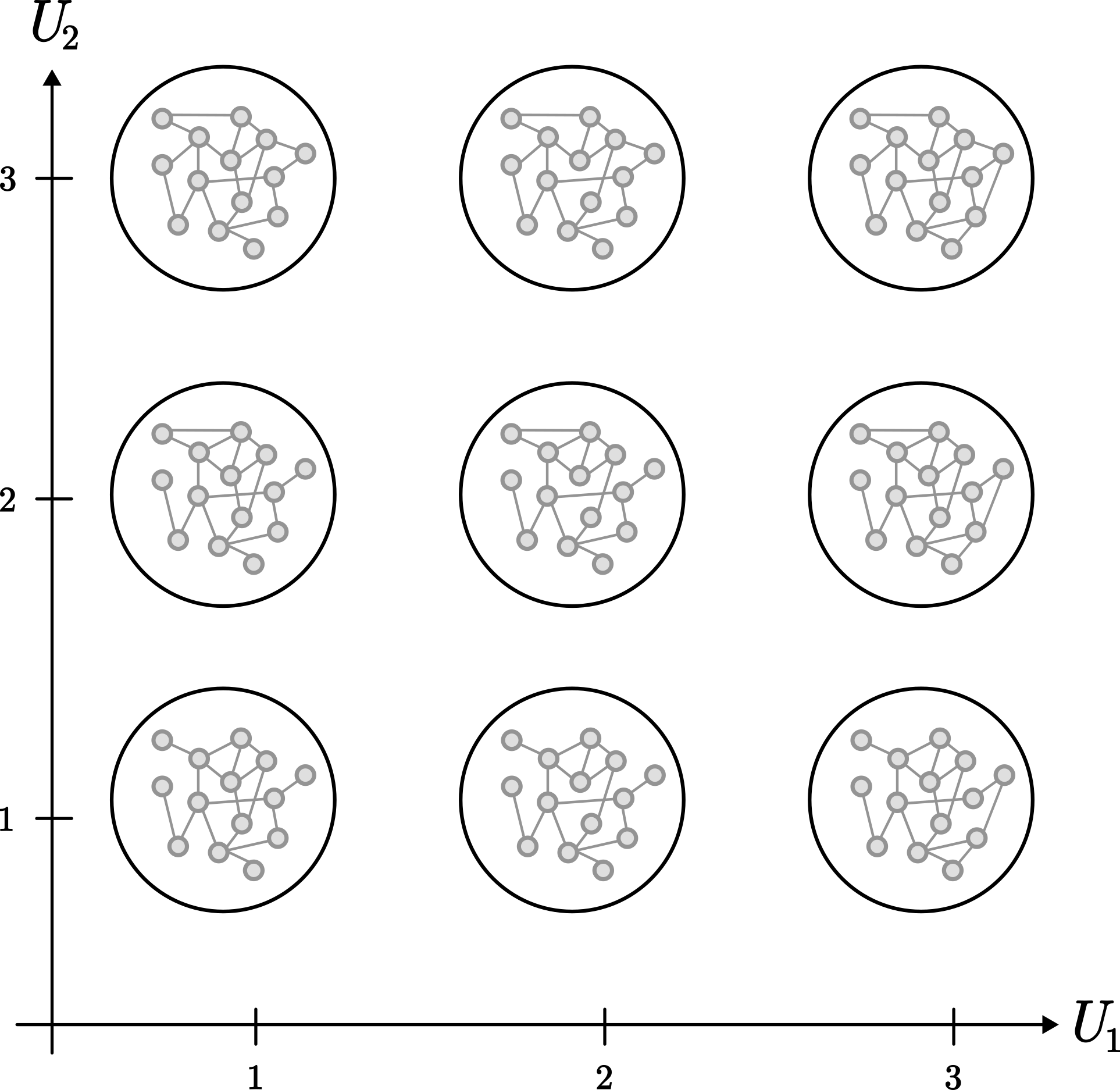}
    \caption{No smoothing, $\b{W}_0$}
    \label{fig:weightMatrixNoSmoothing}
\end{subfigure}
    \hfill
\begin{subfigure}{.3\textwidth}
    \centering
    \includegraphics[width=\textwidth]{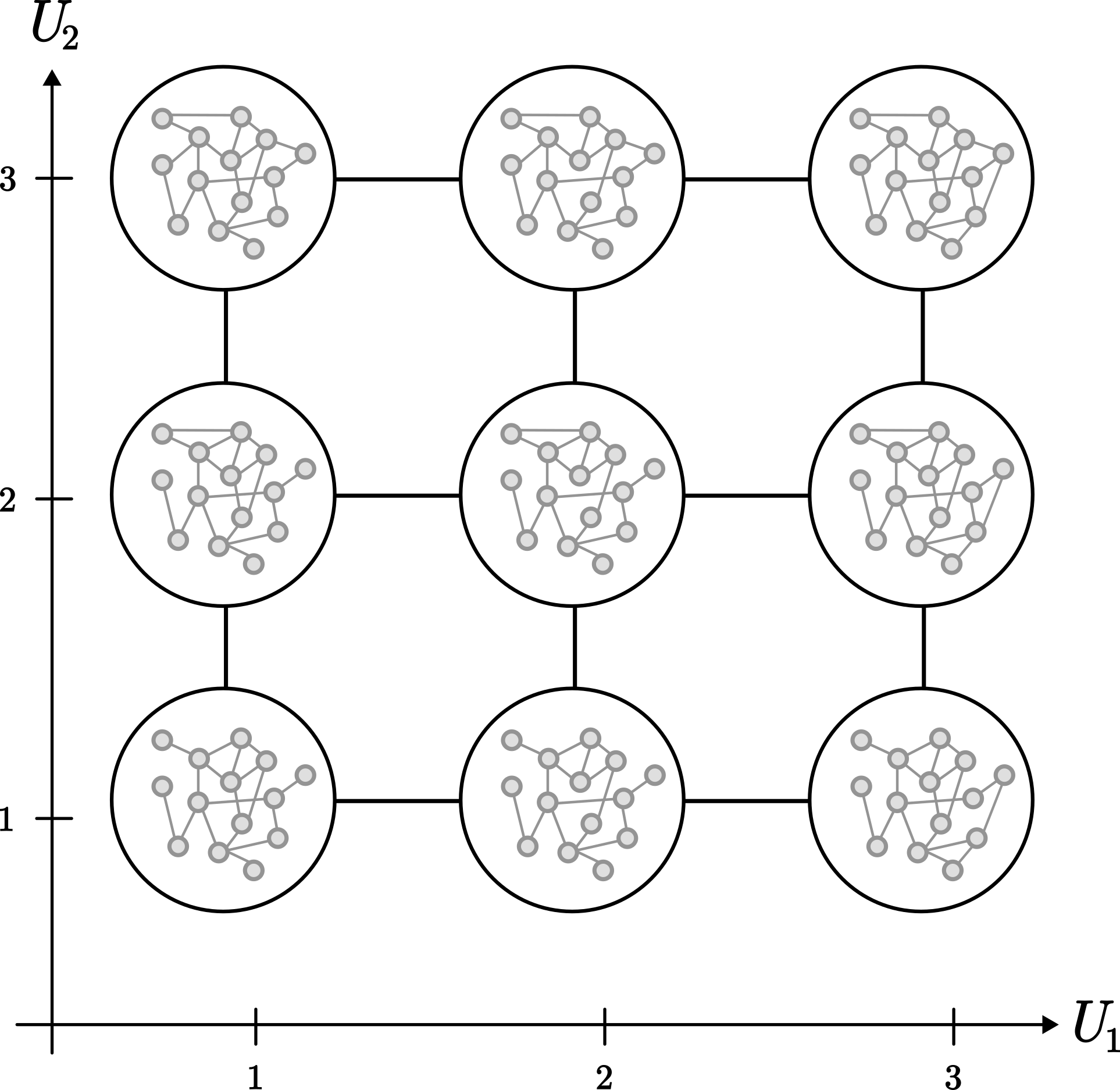}
    \caption{Grid, $\b{W}_\text{grid}$}
    \label{fig:weightMatrixGrid}
\end{subfigure}
    \hfill
\begin{subfigure}{.3\textwidth}
    \centering
    \includegraphics[width=\textwidth]{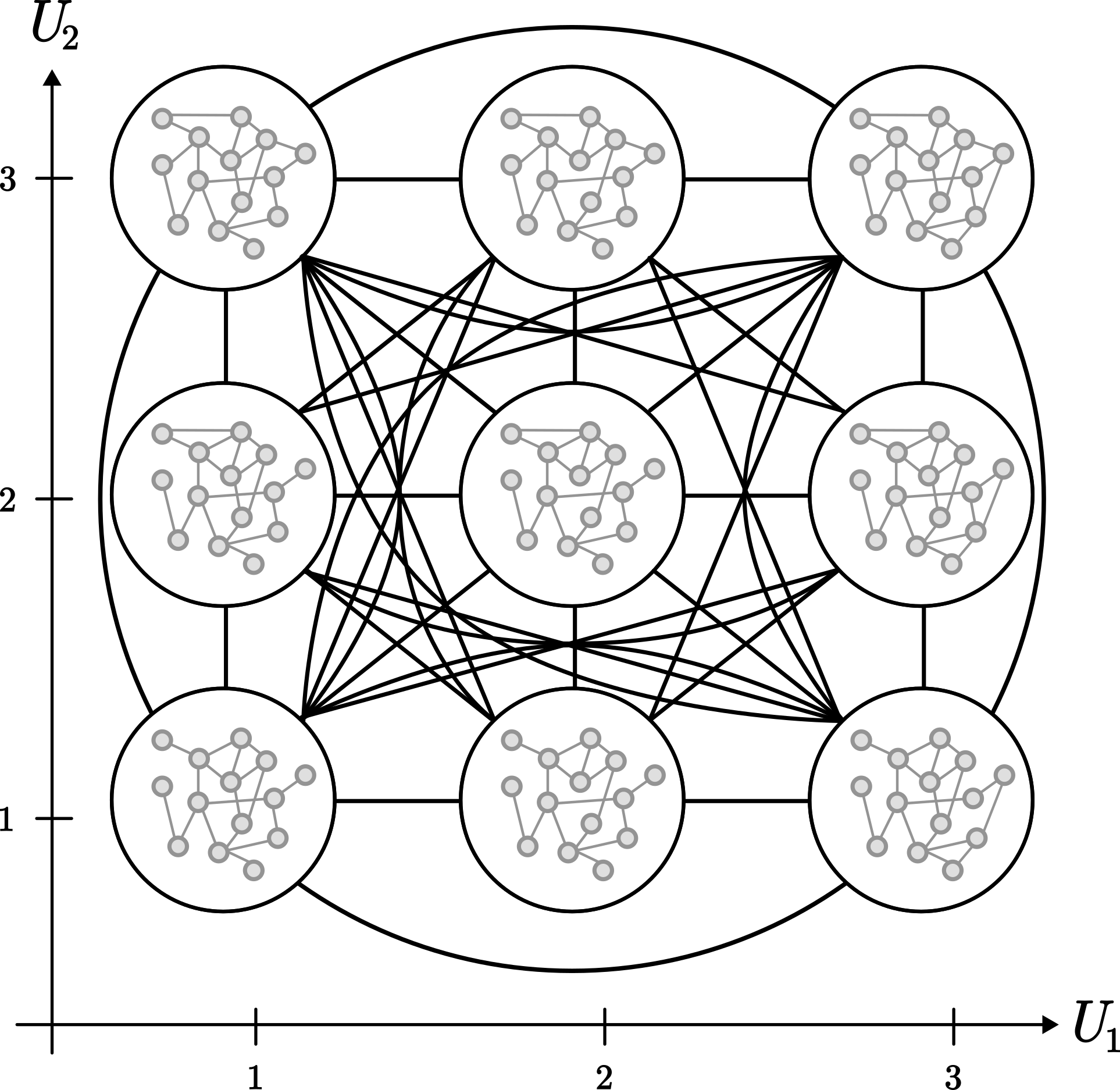}
    \caption{Fully connected, $\b{W}_{\text{full}}$}
    \label{fig:weightMatrixFull}
\end{subfigure}
\RawCaption{\caption{A visualization of the three meta-graphs corresponding to the three weight matrices considered in the simulation study.}
\label{fig:simulationWeightMatrices}}
\end{figure}

\section{Real Data Analysis} \label{sec:realDataAnalysis}

High doses of ionizing radiation (2 Gray [Gy]) used in childhood cancer treatment are known risk factors for acute myeloid leukemia in children and for second primary neoplasms later in life. However, the effects of low doses of radiation ($\leq .05$ Gy), typically used in medical diagnostics, on the incidence of childhood cancer are less understood.

Our case study uses data from the German `Cancer in childhood and
molecular-epidemiology' study (KiKme), which was collected between 2013 and 2019 from 591 former childhood cancer patients registered in the German Childhood Cancer Registry and cancer-free controls \citep{Marron2021}. Within a sub-sample of carefully matched 156 participants, a radiation experiment was conducted where individual reactions of long non-coding ribonucleic acids (lncRNAs) to different radiation exposures in normal somatic cells of the same person were investigated. Primary fibroblasts were donated from long-term childhood cancer survivors who had a first primary cancer during childhood (N1, $n = 52$), with at least one second primary neoplasm (N2+, $n = 52$), and tumor-free controls (N0, $n= 52$) \citep{Grandt2022}. Cultured fibroblasts of each donor were exposed to 0 Gy (sham-irradiation), .05 Gy, and 2 Gy X-rays to investigate cellular reaction to ionizing radiation. 

We used $p = 191$ genes that are annotated with the GO-term `response to ionizing radiation' (GO:0010212). All gene expressions were standardized with mean 0 and a standard deviation of 1 within each class. We estimated all nine graphs based on the various combinations of the levels of the external covariates \emph{group} (N0, N1, N2+) and \emph{radiation} (0, .05 and 2 Gy). 

The tuning parameter to control the sparsity level was set to $\gamma_1 = 10^{-5}$ to achieve very sparse networks and $\gamma_2 = 10^{-6}$ to enforce smoothness. We assumed a regular grid $\b{W}_\text{grid}$, see eq.~\eqref{eq:weightMatrixGrid}, and a grid with weights accordingly to the radiation doses:

\begin{equation*}
   \b{W}_\text{rad} =  \begin{bmatrix}
    0     & .025 & 0    & .5   & 0    & 0    & 0    & 0    & 0 \\ 
    .025  & 0    & .975 & 0    & .5   & 0    & 0    & 0    & 0 \\ 
    0     & .975 & 0    & 0    & 0    & .5   & 0    & 0    & 0 \\ 
    .5    & 0    & 0    & 0    & .025 & 0    & .5   & 0    & 0 \\ 
    0     & .5   & 0    & .025 & 0    & .975 & 0    & .5   & 0 \\ 
    0     & 0    & .5   & 0    & .975 & 0    & 0    & 0    & .5 \\ 
    0     & 0    & 0    & .5   & 0    & 0    & 0    & .025 & 0 \\ 
    0     & 0    & 0    & 0    & .5   & 0    & .025 & 0    & .975 \\ 
    0     & 0    & 0    & 0    & 0    & .5   & 0    & .975 & 0 \\ 
\end{bmatrix}
\end{equation*}

\section{Results} \label{sec:results}

In this section, we present the results from both the simulation and the case study.

\subsection{Simulation Study}

We consider 12 different parameter settings and repeat the simulation 20 times for each setting,  see Section~\ref{sec:simulationSetUp}. We evaluate the $F_1$-score using the AIC and BIC criteria for selecting the tuning parameters $\gamma_1$ and $\gamma_2$, see eq.~\eqref{eq:definitionF1scoreUsingAICorBIC}. Additionally, we assess the performance in an `oracle' scenario where the true graphs are known, see eq.~\eqref{eq:F1Oracle}. 

\begin{figure}
    \centering
    \includegraphics[width=\textwidth]{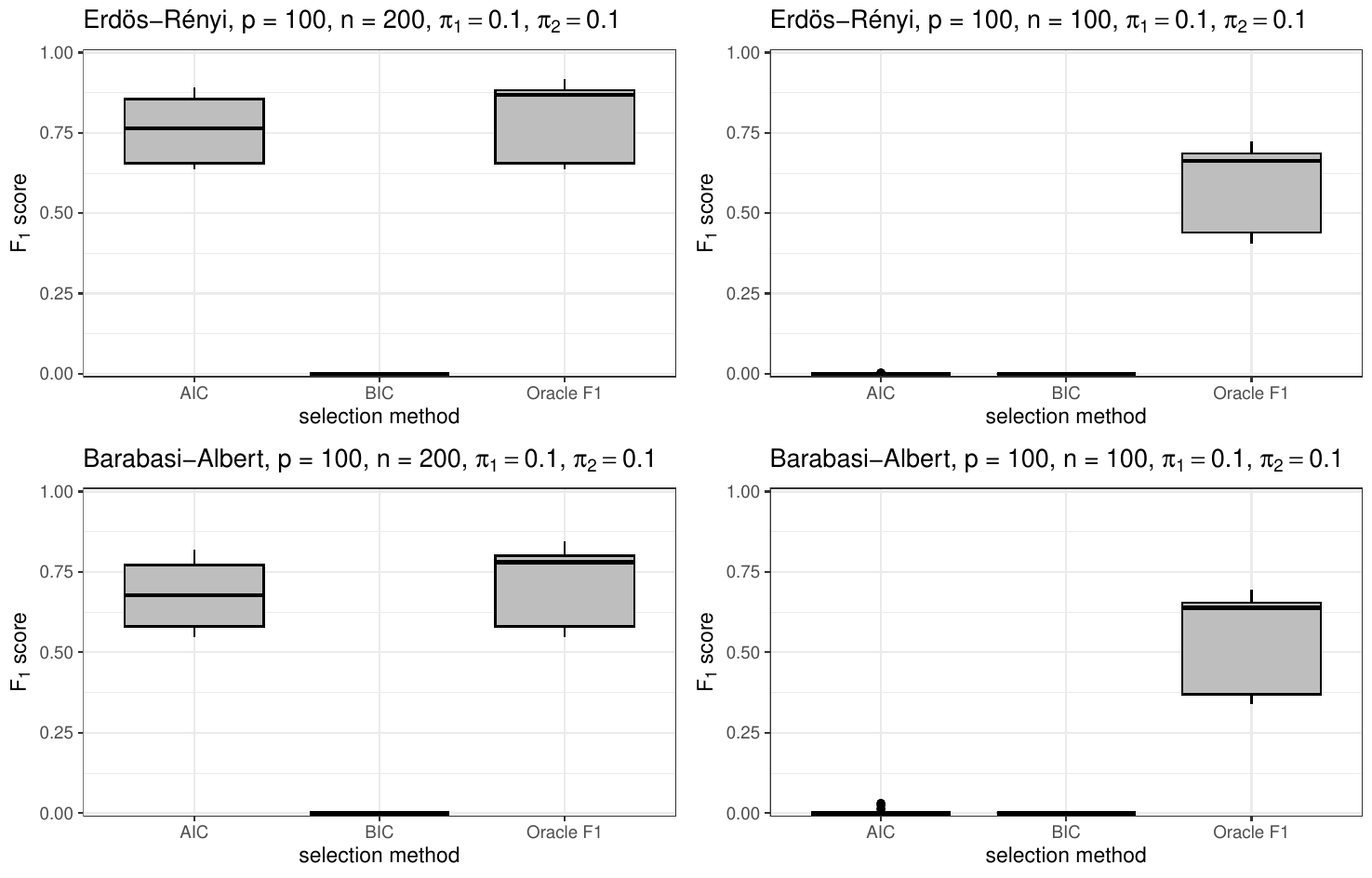}
    \caption{The $F_1$-scores when the tuning parameters are selected based on the AIC, BIC, or an oracle. The top row shows the results for the ER graphs, while the bottom row shows the results for the BA graphs. The left column corresponds to $n = 100$ observations, and the right column corresponds to $n = 200$ observations. The parameters $\pi_1 = \pi_2 = .1$ for all plots.}
    \label{fig:simulationResultsAICBIC}
\end{figure}

\begin{figure}
    \centering
    \includegraphics[width=\textwidth]{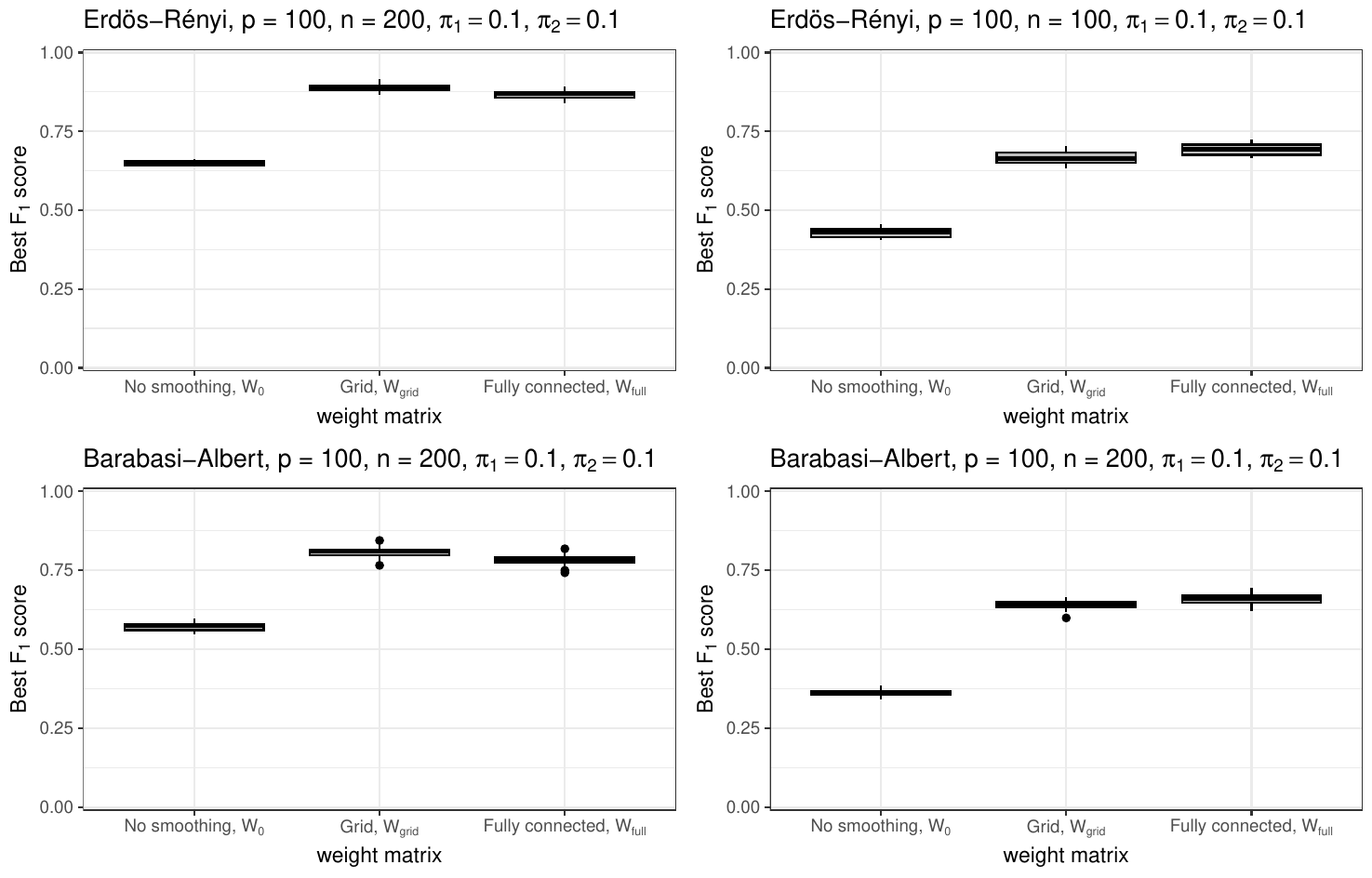}
    \caption{The $F_1$-scores when the tuning parameters $\gamma_1$ and $\gamma_2$ are selected using an oracle, with the CVN model using either $\mathbf{W}_0$ (no smoothing), $\mathbf{W}_\text{grid}$, or $\mathbf{W}_\text{full}$ as the weight matrix. The top row displays results for the ER graphs, and the bottom row shows results for the BA graphs. The left column corresponds to $n = 100$ observations, and the right column corresponds to $n = 200$ observations. The parameters $\pi_1 = \pi_2 = .1$ for all plots.}
    \label{fig:simulationResultsWeightMatrices}
\end{figure}

\begin{figure}
    \centering
    \includegraphics[width=\textwidth]{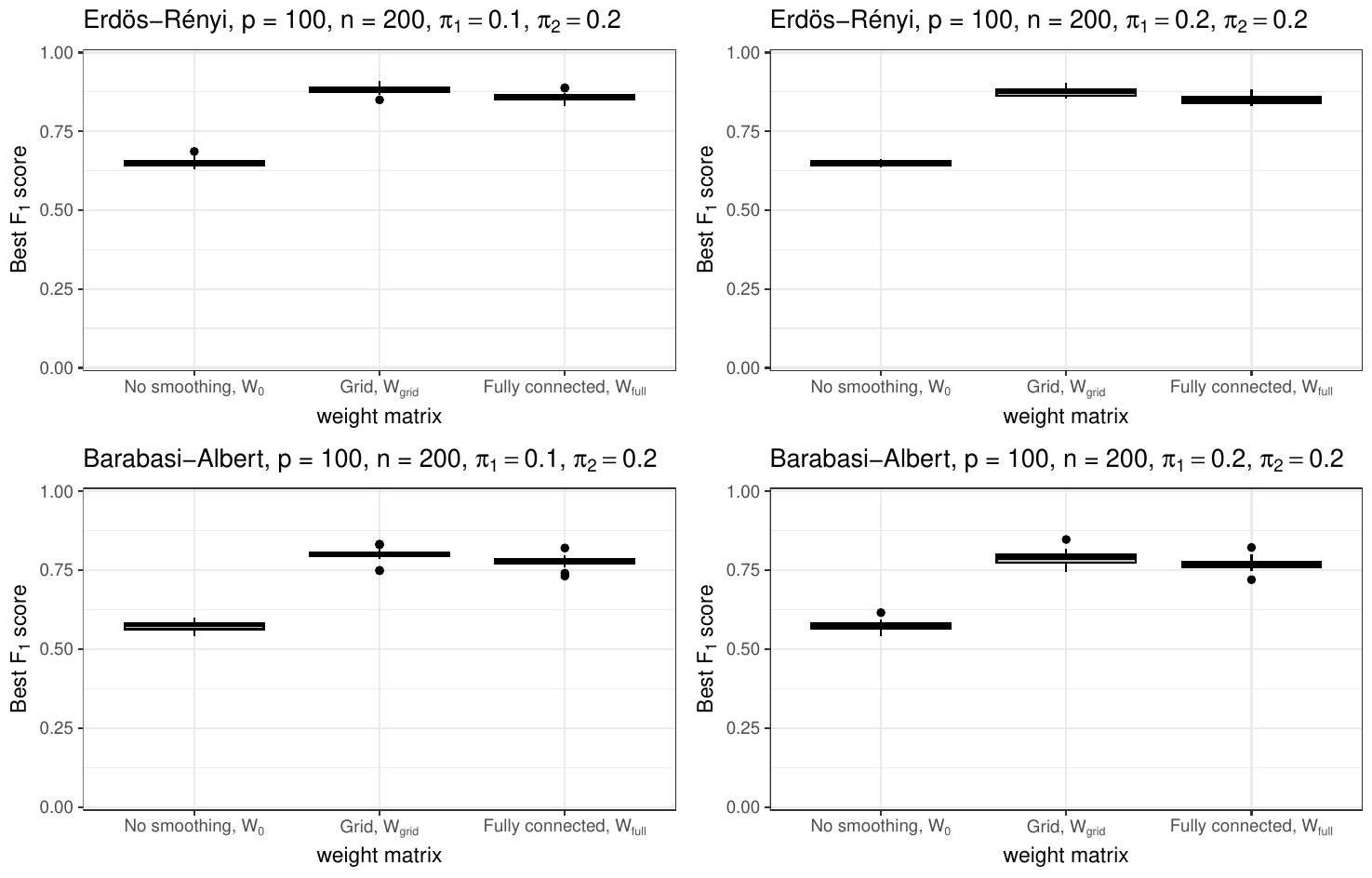}
    \caption{The $F_1$-scores when the tuning parameters $\gamma_1$ and $\gamma_2$ are selected using an oracle, with the CVN model using either $\mathbf{W}_0$ (no smoothing), $\mathbf{W}_\text{grid}$, or $\mathbf{W}_\text{full}$ as the weight matrix. The top row displays results for the ER graphs, and the bottom row shows results for the BA graphs. The number of observations is $n = 200$ for all plots. In the left column, $\pi_1 = .1$ and $\pi_2 = .2$, while in the right column, $\pi_1 = .2$ and $\pi_2 = .2$.}
    \label{fig:simulationResultsChangesInXY}
\end{figure}

Figures~\ref{fig:simulationResultsAICBIC}, \ref{fig:simulationResultsWeightMatrices}, and \ref{fig:simulationResultsChangesInXY} show results from the simulation study. These plots share a common structure: the top row displays results for the ER graphs, while the bottom row shows results for the BA graphs. Additionally, the left column corresponds to the setting with $n = 200$ observations, and the right column corresponds to the setting with $n = 100$ observations.

Figure~\ref{fig:simulationResultsAICBIC} shows the  $F_1$-scores obtained using three different methods for selecting the `optimal' tuning parameters: AIC, BIC, and the oracle. These results pertain to the setting where $\pi_1 = \pi_2 = .1$; similar patterns are observed for other values of $\pi_1$ and $\pi_2$. The AIC performs reasonably well when $n = 200$, but extremely poorly when $n = 100$. The BIC performs poorly in both cases, since it tends to be too stringent, resulting in very sparse graphs. However, if the optimal values were known, one could achieve rather good $F_1$-scores. This is shown by the oracle which performs best since it considers the true underlying structure of the CVN.
This at least shows that, in principle, the CVN can be estimated rather well.
Due to the poor performance of both information criteria, we consider the oracle $F_1$-score for the rest of this section.  

Figure~\ref{fig:simulationResultsWeightMatrices} shows the best $F_1$-score achievable by selecting the `optimal' $\gamma_1$ and $\gamma_2$ for different weight matrices. Recall that, as described in Section~\ref{sec:simulation}, the grid weight matrix, $\mathbf{W}_\text{grid}$, aligns best with the data-generating process. As expected, using the grid weight matrix yields the best results when $n = 200$. However, with fewer observations (right column), the fully connected weight matrix, $\mathbf{W}_\text{full}$, performs slightly better. This might be due to the fact that smoothing all nine graphs simultaneously increases the `evidence' for the existence of edges.

Figure~\ref{fig:simulationResultsChangesInXY} shows the best $F_1$-score when the extent to which the graphs change in the $U_1$ and $U_2$ directions varies. Whether 10\% or 20\% of the expected number of edges change in these directions has little impact on the algorithm's performance. This pattern is consistent across other simulation settings as well. 

All the simulation results can be found and explored interactively at \url{cvn.bips.eu}.

\subsection{Case Study}

\begin{figure}[htbp]
    \centering
    \begin{subfigure}{0.3\textwidth}
        \centering
        \includegraphics[width=\textwidth]{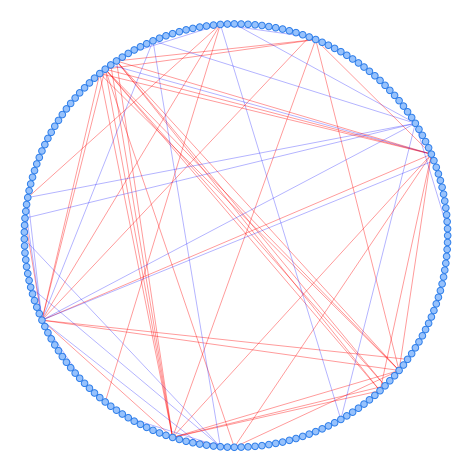}
        \caption{N2+, 0 Gy}
    \end{subfigure}
    \begin{subfigure}{0.3\textwidth}
        \centering
        \includegraphics[width=\textwidth]{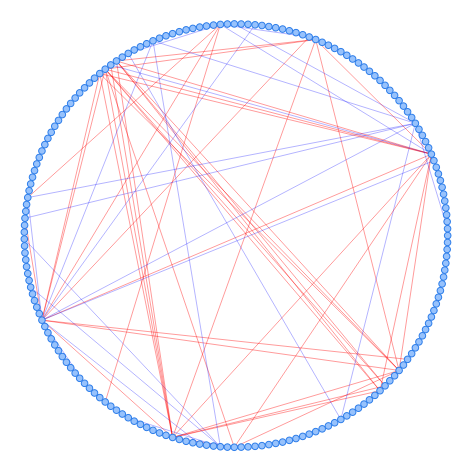}
        \caption{N2+, .05 Gy}
    \end{subfigure}
    \begin{subfigure}{0.3\textwidth}
        \centering
        \includegraphics[width=\textwidth]{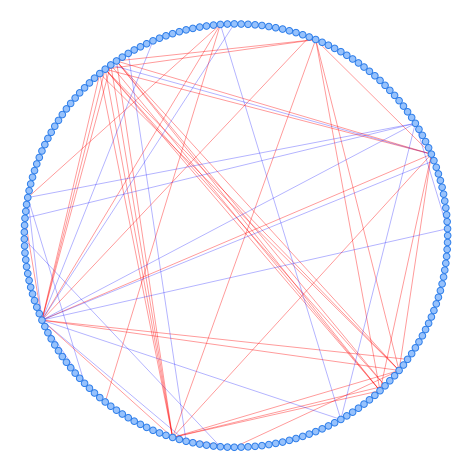}
        \caption{N2+, 2 Gy}
    \end{subfigure}
    
    \begin{subfigure}{0.3\textwidth}
        \centering
        \includegraphics[width=\textwidth]{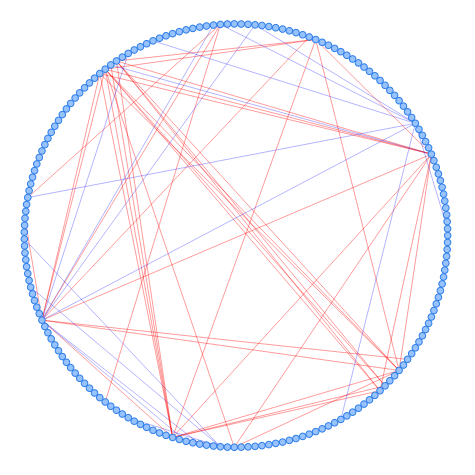}
        \caption{N1, 0 Gy}
    \end{subfigure}
    \begin{subfigure}{0.3\textwidth}
        \centering
        \includegraphics[width=\textwidth]{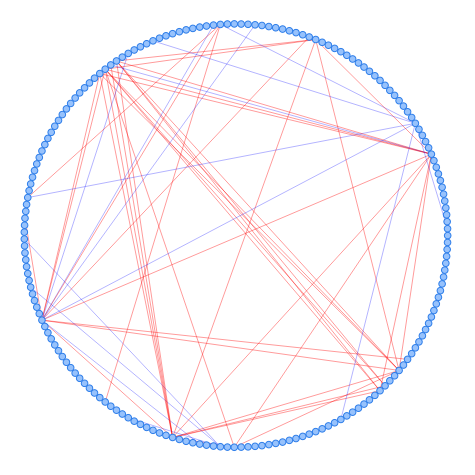}
        \caption{N1, .05 Gy}
    \end{subfigure}
    \begin{subfigure}{0.3\textwidth}
        \centering
        \includegraphics[width=\textwidth]{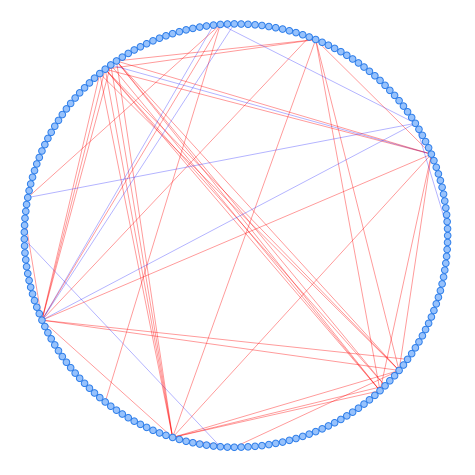}
        \caption{N1, 2 Gy}
    \end{subfigure}
    
    \begin{subfigure}{0.3\textwidth}
        \centering
        \includegraphics[width=\textwidth]{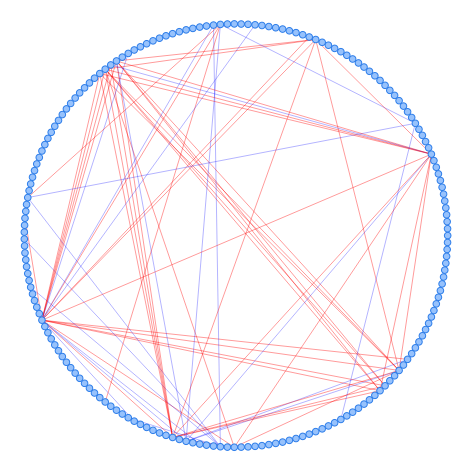}
        \caption{N0, 0 Gy}
    \end{subfigure}
    \begin{subfigure}{0.3\textwidth}
        \centering
        \includegraphics[width=\textwidth]{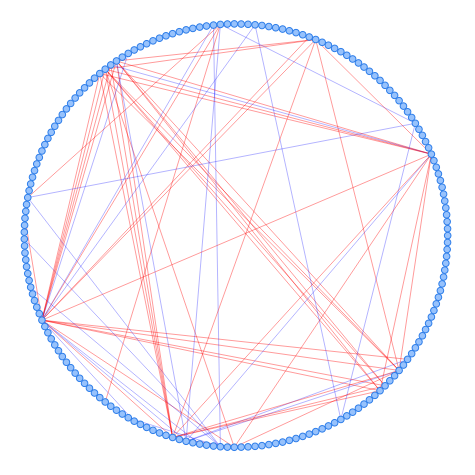}
        \caption{N0, .05 Gy}
    \end{subfigure}
    \begin{subfigure}{0.3\textwidth}
        \centering
        \includegraphics[width=\textwidth]{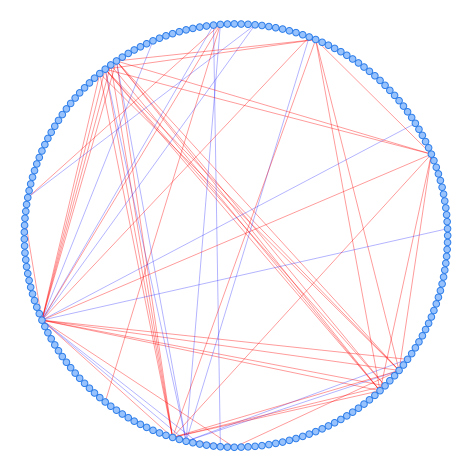}
        \caption{N0, 2 Gy}
    \end{subfigure}
    
    \caption{Estimated graphs based on $\b{W}_\text{rad}$, $\gamma_1 = 5 \times 10^{-5}$ and $\gamma_2 = 5 \times 10^{-6}$. Blue lines denote $42$ edges present in \emph{all} nine graphs.}
    \label{fig:kikme}
\end{figure}

Figure~\ref{fig:kikme} shows the nine estimated graphs based on the KiKme dataset with $\gamma_1 = 5 \times 10^{-5}$ and $\gamma_2 = 5 \times 10^{-6}$ and the weight matrix  $\b{W}_\text{rad}$. The nodes, corresponding to the $p = 191$ gene expressions, are positioned on a circle, with lines indicating the edges. Red lines denote $42$ edges present in \emph{all} nine graphs, representing the `core graph' unaffected by external covariates, while blue lines indicate edges present in only a subset of the graphs, i.e., the edges that are influenced by the external covariates. 

Here, the focus is not just on the graphs themselves, but also on the level of dis)similarity between them. Here, we consider the Hamming distances between graphs. Let $\widehat{\b{A}}_{ij}\left(\gamma_1, \gamma_2\right)$ and $\widehat{\b{A}}_{kl}\left(\gamma_1, \gamma_2\right)$ denote the estimated adjacency matrices when $(U_1 = i, U_2 = j)$ and $(U_1 = k, U_2 = l)$, respectively, and the tuning parameters are $\gamma_1$ and $\gamma_2$. The Hamming distance between these two graphs is defined as the number of edges that is present in one graph but absent in the other: 
\begin{equation*}
	\text{Hamming distance}\left(\gamma_1, \gamma_2\right) = \sum_{s < t} \mathds{1}\left( \widehat{a}_{st}^{(ij)}\left(\gamma_1, \gamma_2\right) \neq \widehat{a}_{st}^{(kl)}\left(\gamma_1, \gamma_2\right) \right).
\end{equation*}
This distance reflects the number of edges that must be added and/or removed to transform one graph into another. 

\begin{figure}[h!]
 \centering
  \includegraphics[width=\textwidth]{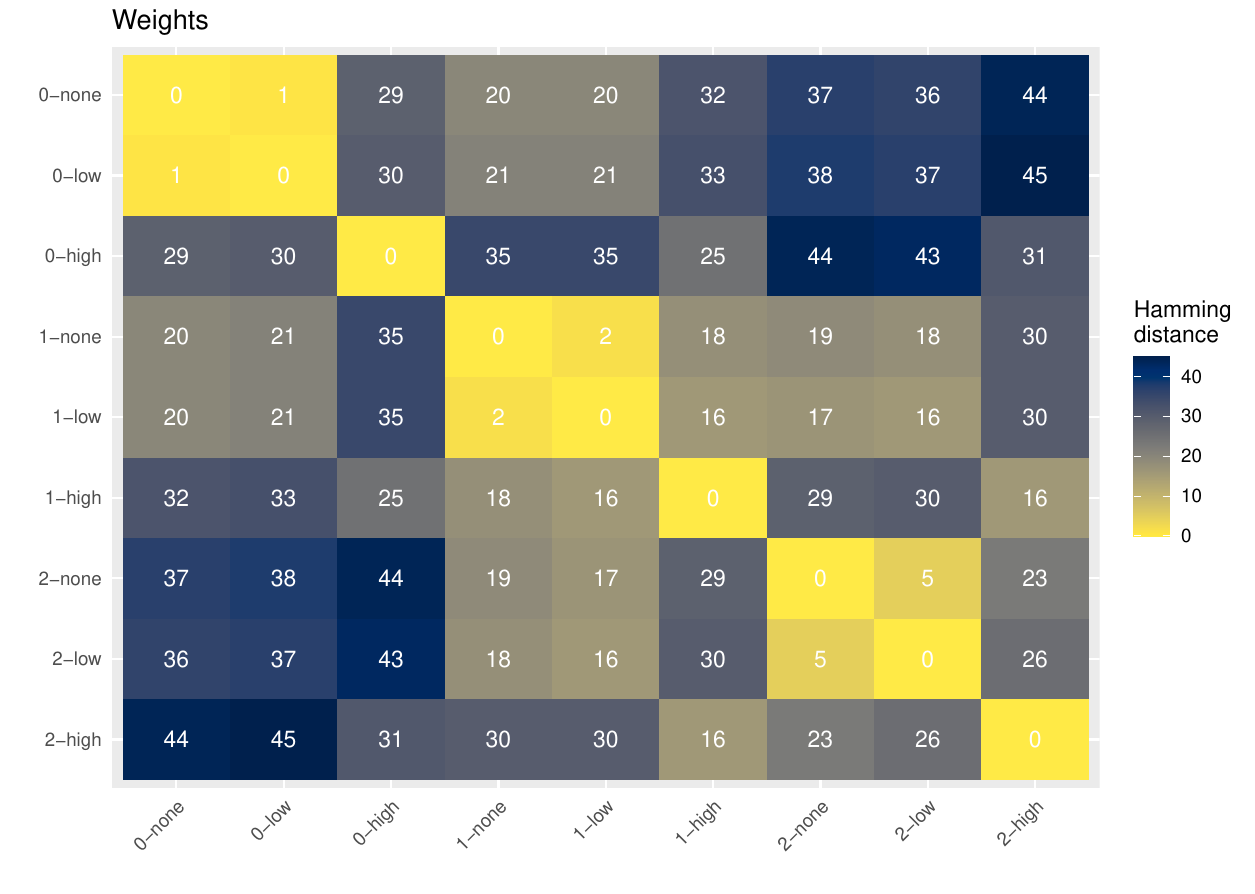}
 \caption{The Hamming distances between the nine graphs of a CVN with $\b{W}_\text{rad}$ based on the KiKme data set when $\gamma_1 = 5 \times 10^{-5}$ and $\gamma_2 = 5 \times 10^{-6}$. The rows/columns are grouped together based on the cancer group the individuals belong to.}
 \label{fig:heatmap}
\end{figure}

Figure~\ref{fig:heatmap} shows a heatmap of the Hamming distances between the nine graphs. 
The first three rows/columns represent the N0 control group, the fourth through sixth rows/columns represent the N1 results, and the last three rows/columns represent the N2+ results. Within these groups, the first row/column corresponds to no radiation, the second to .05 Gy, and the third to the highest dose of 2 Gy. The color gradient represents the Hamming distances. 
Within a group, there is almost no difference between no radiation and low radiation. However, these networks are clearly different from those with the maximum radiation dose. The networks between cancer groups within an intensity of irradiation are clearly different, regardless of the radiation dose. 

\begin{table}[h!]
\centering
{\footnotesize
\begin{tabular}{ccllllll}
  \toprule
   \textbf{Group} & \textbf{Gy} &
   \rotatebox{90}{\textbf{Number of nodes}} &%
   \rotatebox{90}{\textbf{Number of edges}} &%
   \rotatebox{90}{\textbf{Graph density}} &%
   \rotatebox{90}{\textbf{Avg. shortest path length}} &%
   \rotatebox{90}{\textbf{Avg. Degree}} &%
   \rotatebox{90}{\textbf{Max. Degree and nodes}} \\ 
  \midrule
  \textbf{N0} & \textbf{0} & 31 & 75 & .16 & 2.38 & 4.84 & 21 \\ 
  \textbf{N0} & \textbf{.05} & 31 & 76 & .16 & 2.33 & 4.9 & 21 \\ 
  \textbf{N0} & \textbf{2} & 27 & 68 & .19 & 2.15 & 5.04 & 21 \\ 
  \hline
  \textbf{N1} & \textbf{0} & 31 & 63 & .14 & 2.33 & 4.06 & 15 \\ 
  \textbf{N1} & \textbf{.05} & 30 & 61 & .14 & 2.34 & 4.07 & 14 \\ 
  \textbf{N1} & \textbf{2} & 25 & 57 & .19 & 2.24 & 4.56 & 13 \\ 
  \hline
  \textbf{N2+} & \textbf{0} & 32 & 66 & .13 & 2.69 & 4.12 & 16 \\ 
  \textbf{N2+} & \textbf{.05} & 32 & 67 & .14 & 2.65 & 4.19 & 16 \\ 
  \textbf{N2+} & \textbf{2} & 29 & 65 & .16 & 2.16 & 4.48 & 19 \\ 
   \bottomrule
\end{tabular}
\caption{Characteristics of the 9 estimated subgraphs. Nodes in each subgraph have at least one adjacent node. Singletons were removed from the original graph with $p = 191$ nodes.}
\label{tab:kikme1}
}
\end{table}

Table~\ref{tab:kikme1} demonstrates that within each CVN subgraph, only a subset of nodes are interconnected. There are only few differences within a group between radiation doses of 0 and .05 Gy, which may be due to random variation. Notably, the size of the subsets decreases with increasing radiation dose, whereas the density and the average node degree in each subgraph increases, implying a higher level of inter-connectivity among the remaining genes. The TP53 gene consistently exhibits the highest degree centrality, indicating its importance in the network regardless of the cancer group or radiation dosage. The TP53 gene encodes the tumor suppressor protein p53 that regulates DNA repair, apoptosis, and cell cycle control. It plays a critical role in the prevention of cancer, and mutations in this gene are common in human tumors.

\begin{table}
\centering
{\tiny
\rotatebox{90}{%
\begin{tabular}{llllrrrrrrrrr}
  \toprule
\textbf{Gene} & \textbf{Chr} & \textbf{Description} & \textbf{Found in} & \rotatebox{90}{N00} & \rotatebox{90}{N01} & \rotatebox{90}{N02} & \rotatebox{90}{N10} & \rotatebox{90}{N11} & \rotatebox{90}{N12} & \rotatebox{90}{N20} & \rotatebox{90}{N21} & \rotatebox{90}{N22} \\ 
  \midrule
ATG4A & X & Autophagy Related 4a Cysteine Peptidase & N2+ & 0 & 0 & 0 & 0 & 0 & 0 & 0 & 0 & 1 \\ 
  RAD9B & 12 & Rad9 Checkpoint Clamp Component B & N2 & 0 & 0 & 0 & 0 & 0 & 0 & 1 & 0 & 0 \\ 
  STX6 & 1 & Syntaxin 6 & N2+ & 0 & 0 & 0 & 0 & 0 & 0 & 1 & 1 & 1 \\ 
  TRIM13 & 13 & Tripartite Motif Containing 13 & N2+ & 0 & 0 & 0 & 0 & 0 & 0 & 1 & 1 & 1 \\ 
  XRCC4 & 5 & X-Ray Repair Cross Complementing 4 & N2+ & 0 & 0 & 0 & 0 & 0 & 0 & 1 & 1 & 1 \\ 
  CES2 & 16 & Carboxylesterase 2 & 2 Gy & 0 & 0 & 0 & 0 & 0 & 1 & 0 & 0 & 1 \\ 
  SESN2 & 1 & Sestrin 2 & 2 Gy & 0 & 0 & 1 & 0 & 0 & 0 & 0 & 0 & 1 \\ 
  FBXO22 & 15 & F-Box Protein 22 & in cancer group(s) and for N0 with 2 Gy & 0 & 0 & 1 & 0 & 0 & 0 & 1 & 1 & 1 \\ 
  XRCC6 & 22 & X-Ray Repair Cross Complementing 6 & in cancer group(s) and for N0 with 2 Gy & 0 & 0 & 1 & 1 & 1 & 1 & 1 & 1 & 1 \\ 
  GRB2 & 17 & Growth Factor Receptor Bound Protein 2 & control group & 1 & 1 & 0 & 0 & 0 & 0 & 0 & 0 & 0 \\ 
  PRKAA1 & 5 & Protein Kinase Amp-Activated Catalytic Subunit Alpha 1 & control group & 1 & 1 & 0 & 0 & 0 & 0 & 0 & 0 & 0 \\ 
  TAF3 & 10 & Tata-Box Binding Protein Associated Factor 3 & in N0 and N1 with $<$= .05 Gy & 1 & 1 & 0 & 1 & 1 & 0 & 0 & 0 & 0 \\ 
  NOX4 & 11 & Nadph Oxidase 4 & control group & 1 & 1 & 1 & 0 & 0 & 0 & 0 & 0 & 0 \\ 
  BAX & 19 & Bcl2 Associated X, Apoptosis Regulator & control group & 1 & 1 & 1 & 0 & 0 & 0 & 0 & 0 & 0 \\ 
  ASCC3 & 6 & Activating Signal Cointegrator 1 Complex Subunit 3 & control group & 1 & 1 & 1 & 0 & 0 & 0 & 0 & 0 & 0 \\ 
  TP53INP1 & 8 & Tumor Protein P53 Inducible Nuclear Protein 1 & control group & 1 & 1 & 1 & 0 & 0 & 0 & 0 & 0 & 0 \\ 
  SPIDR & 8 & Scaffold Protein Involved In Dna Repair & 0-2 Gy in control group, unclear pattern for cancer groups & 1 & 1 & 1 & 0 & 0 & 1 & 0 & 0 & 1 \\ 
  DNMT3A & 2 & Dna Methyltransferase 3 Alpha & 0-2 Gy in control group, unclear pattern for cancer groups & 1 & 1 & 1 & 1 & 0 & 0 & 0 & 0 & 1 \\ 
  NABP1 & 2 & Nucleic Acid Binding Protein 1 & 0-2 Gy in control group, unclear pattern for cancer groups & 1 & 1 & 1 & 1 & 1 & 0 & 0 & 1 & 0 \\ 
  GADD45A & 1 & Growth Arrest And Dna Damage Inducible Alpha & 0-2 Gy in control group, unclear pattern for cancer groups & 1 & 1 & 1 & 1 & 1 & 0 & 1 & 1 & 0 \\ 
  RHNO1 & 12 & Rad9-Hus1-Rad1 Interacting Nuclear Orphan 1 & 0-2 Gy in control group, unclear pattern for cancer groups & 1 & 1 & 1 & 1 & 1 & 1 & 0 & 0 & 0 \\ 
  RAD9A & 11 & Rad9 Checkpoint Clamp Component A & in all graphs & 1 & 1 & 1 & 1 & 1 & 1 & 1 & 1 & 1 \\ 
  HMGA2 & 12 & High Mobility Group At-Hook 2 & in all graphs & 1 & 1 & 1 & 1 & 1 & 1 & 1 & 1 & 1 \\ 
  BLM & 15 & Blm Recq Like Helicase & in all graphs & 1 & 1 & 1 & 1 & 1 & 1 & 1 & 1 & 1 \\ 
  XRCC5 & 2 & X-Ray Repair Cross Complementing 5 & in all graphs & 1 & 1 & 1 & 1 & 1 & 1 & 1 & 1 & 1 \\ 
  CASP3 & 4 & Caspase 3 & in all graphs & 1 & 1 & 1 & 1 & 1 & 1 & 1 & 1 & 1 \\ 
  ANKRA2 & 5 & Ankyrin Repeat Family A Member 2 & in all graphs & 1 & 1 & 1 & 1 & 1 & 1 & 1 & 1 & 1 \\ 
  SESN1 & 6 & Sestrin 1 & in all graphs & 1 & 1 & 1 & 1 & 1 & 1 & 1 & 1 & 1 \\ 
  TNFRSF10B & 8 & Tnf Receptor Superfamily Member 10b & in all graphs & 1 & 1 & 1 & 1 & 1 & 1 & 1 & 1 & 1 \\ 
  ACER2 & 9 & Alkaline Ceramidase 2 & in all graphs & 1 & 1 & 1 & 1 & 1 & 1 & 1 & 1 & 1 \\ 
   \bottomrule
\end{tabular}
}}
\caption{Genes ajdacent to TP53 in the estimated CVN.}
\label{tab:kikme2}
\end{table}

Table~\ref{tab:kikme2} shows all adjacent genes of TP53 in each subgraph, nine of them being shared by all of them. Interestingly, TP53 is only associated with BAX in the control group for all radiation doses. The expression of BAX is upregulated in the presence of the tumor suppressor p53 following significant cellular damage, leading to the initiation of apoptotic (programmed) cell death. This is beneficial for cancer prevention. If there is a mutation in either TP53 or BAX, the  process of TP53 initiating BAX may be disrupted and contribute to cancer development.

\section{Conclusions \& Discussion} \label{sec:conclusions}

We introduced a new Gaussian graphical model called the covariate-varying network (CVN). This model allows the graph structure to change with multiple discrete external covariates, e.g., time and groups. To achieve this, we applied a LASSO penalty to the entries of the precision matrix to introduce sparsity. Additionally, we enforce similarities, or `smoothness', between the different graphs by penalizing the absolute differences between them. We modeled smoothness using the concept of a meta graph with weighted edges, see Section~\ref{sec:cvn}. This approach is the first to address the problem of a graph changing with multiple external covariates. Many previously existing graphical models are special cases within the CVN graphical model class, see Section~\ref{sec:specialcases}. 

We solved the resulting maximum likelihood optimization problem associated with the CVN model, see eq.~\eqref{eq:cvn}, using an Alternating Direction Method of Multipliers (ADMM), see Section~\ref{sec:algorithm}. In the second update step of the algorithm (Section~\ref{sec:updateZ}), we discovered that the problem could be subdivided into individual optimization problems, one for each edge. This subproblem is equivalent to a weighted Fused LASSO Signal Approximator (wFLSA), for which we developed a specialized algorithm previously, see \citet{dijkstra2024spinoff}.

We determined that the complexity of our algorithm is cubic with respect to the number of variables ($p^3$) and squared with respect to the number of graphs ($m^2$), see Section~\ref{sec:computationalComplexity}. This is a downside, as the algorithm becomes computationally expensive and challenging to run for very large instances. The primary bottleneck is the second update step, where the wFLSA must be solved for each of the $\binom{p}{2}$ edges. 

We proposed a different type of tuning parameter parameterization as is common for graphical models, see Section~\ref{sec:alternativeTuningParameterization}. Instead of using tuning parameters that penalize the $L_1$-norm of the precision matrix ($\lambda_1$) and the absolute differences between precision matrices ($\lambda_2$), our proposed parameters penalize individual entries of a precision matrix ($\gamma_1$) and individual differences between individual edges ($\gamma_2$). The advantage of this approach is that once appropriate values for $\gamma_1$ and $\gamma_2$ are found, they are likely to change only slightly when the dataset is extended with additional variables or graphs, i.e., when $p$ and $m$ change. This stability would not be the case for $\lambda_1$ and $\lambda_2$. Additionally, this parameterization is not only useful for the CVN model but could also benefit other LASSO-type problems.

We conducted a simulation study (see Sections~\ref{sec:simulation} and \ref{sec:simulationSetUp}) using two types of graphs: Erdős–Rényi and Barabasi–Albert. These graphs were influenced by two external covariates, resulting in varied changes to the graph structure based on each covariate. We found that the estimation of the CVN was generally successful across different scenarios. However, selecting the tuning parameters poses significant challenges. 
The extent to which the graphs change with the external covariates did not appear to impact performance. The authors anticipate that if higher percentages were considered, one would see a decline in performance.
It is important to note that comparing this method with others is difficult since, to the best of our knowledge, there are currently no other methods capable of handling the CVN graphical model.

To assess its applicability, we applied our method to a real dataset from the KiKme study (see Section~\ref{sec:realDataAnalysis}). We showed that the method is able to recover differences and similarities of gene expression network structures that are specific to  three different case/control groups and three different radiation dosages. This case study demonstrated that CVN is capable of estimating plausible networks in a high-dimensional setting where the overall number of samples ($n = 156$) is smaller than the number of nodes ($p = 191$). The number of samples for each subgraph was therefore much smaller.

Moving forward, there are several avenues for future research. First, extending the method to handle continuous covariates would be a promising direction, broadening its applicability across different datasets. Currently, dealing with continuous external covariates necessitates data discretization. Second, based on our simulation study, we observed that the performance of AIC and BIC for tuning parameter selection is notably inadequate. Exploring alternative methods for parameter tuning could potentially improve performance. Lastly, due to the high algorithmic complexity of our method, exploring alternative optimization strategies, including heuristic approaches, could help reduce computational costs and enhance efficiency. 

The algorithm has been implemented as an \texttt{R}~package and is publicly accessible at \url{www.github.com/bips-hb/CVN}. Additionally, the CVN data simulator is available as an \texttt{R}~package at \url{www.github.com/bips-hb/CVNSim}, and the code for the simulation study can be found at \url{www.github.com/bips-hb/CVNStudy}. All results of the simulation study are hosted at \url{cvn.bips.eu}. 

\section*{Acknowledgments}
We gratefully acknowledge the financial support of the German Research Foundation (DFG -- Project FO 1045/2-1).








\bibliography{main}

\begin{thebibliography}{}

\bibitem[Ando, 2007]{ando2007bayesian}
Ando, T. (2007).
\newblock Bayesian predictive information criterion for the evaluation of hierarchical bayesian and empirical bayes models.
\newblock {\em Biometrika}, 94(2):443--458.

\bibitem[Badhwar et~al., 2017]{badhwar2017resting}
Badhwar, A., Tam, A., Dansereau, C., Orban, P., Hoffstaedter, F., and Bellec, P. (2017).
\newblock {Resting-state network dysfunction in Alzheimer's disease: A systematic review and meta-analysis}.
\newblock {\em Alzheimer's \& Dementia: Diagnosis, Assessment \& Disease Monitoring}, 8:73--85.

\bibitem[Banjeree and Ghaoui, 2008]{Banjeree2008}
Banjeree, O. and Ghaoui, L.~E. (2008).
\newblock {Model selection through sparse maximum likelihood estimation for multivariate Gaussian or binary data}.
\newblock {\em Journal of Machine Learning Research}, 9:458--516.

\bibitem[Barabasi and Oltvai, 2004]{barabasi2004network}
Barabasi, A.-L. and Oltvai, Z.~N. (2004).
\newblock {Network biology: Understanding the cell's functional organization}.
\newblock {\em Nature Reviews Genetics}, 5(2):101--113.

\bibitem[Boyd et~al., 2010]{Boyd2010}
Boyd, S., Parikh, N., Chu, E., Peleato, B., and Eckstein, J. (2010).
\newblock Distributed optimization and statistical learning via the alternating direction method of multipliers.
\newblock {\em Foundations and Trends in Machine Learning}, 3:1--122.

\bibitem[Boyd and Vandenberghe, 2004]{Boyd2004}
Boyd, S. and Vandenberghe, L. (2004).
\newblock {\em Convex optimization}.
\newblock Cambridge University Press.

\bibitem[Brent, 2013]{brent2013algorithms}
Brent, R.~P. (2013).
\newblock {\em Algorithms for minimization without derivatives}.
\newblock Courier Corporation.

\bibitem[Brozyna et~al., 2007]{brozyna2007mechanism}
Brozyna, A., Zbytek, B., Granese, J., Carlson, J.~A., Ross, J., and Slominski, A. (2007).
\newblock {Mechanism of UV-related carcinogenesis and its contribution to nevi/melanoma}.
\newblock {\em Expert Review of Dermatology}, 2(4):451--469.

\bibitem[Cai et~al., 2011]{Cai2011}
Cai, T., Liu, W., and Luo, X. (2011).
\newblock A constrained $\ell_1$ minimization approach to sparse precision matrix estimation.
\newblock {\em Journal of the American Statistical Association}, 106:594--607.

\bibitem[Chen and Chen, 2008]{chen2008extended}
Chen, J. and Chen, Z. (2008).
\newblock Extended bayesian information criteria for model selection with large model spaces.
\newblock {\em Biometrika}, 95(3):759--771.

\bibitem[Curtis et~al., 2012]{huttenhower2012}
Curtis, H., Blaser, M.~J., Dirk, G., Kota, K.~C., Rob, K., Liu, B., Wang, L., Sahar, A., White, J.~R., Badger, J.~H., et~al. (2012).
\newblock Structure, function and diversity of the healthy human microbiome.
\newblock {\em Nature}, 486(7402):207--214.

\bibitem[Danaher et~al., 2014]{Danaher2014}
Danaher, P., Wang, P., and Witten, D.~M. (2014).
\newblock The joint graphical lasso for inverse covariance estimation across multiple classes.
\newblock {\em Journal of the Royal Statistical Society. Series B: Statistical Methodology}, 76:373--397.

\bibitem[de~la Cuesta-Zuluaga et~al., 2018]{de2018gut}
de~la Cuesta-Zuluaga, J., Corrales-Agudelo, V., Vel{\'a}squez-Mej{\'\i}a, E.~P., Carmona, J.~A., Abad, J.~M., and Escobar, J.~S. (2018).
\newblock Gut microbiota is associated with obesity and cardiometabolic disease in a population in the midst of westernization.
\newblock {\em Scientific Reports}, 8(1):11356.

\bibitem[de~Vos and de~Vos, 2012]{deVos2012}
de~Vos, W.~M. and de~Vos, E.~A. (2012).
\newblock {{Role of the intestinal microbiome in health and disease: From correlation to causation}}.
\newblock {\em Nutrition Reviews}, 70:S45--S56.

\bibitem[Dijkstra et~al., 2024]{dijkstra2024spinoff}
Dijkstra, L., Hanke, M., Koenen, N., and Foraita, R. (2024).
\newblock An alternating direction method of multipliers algorithm for the weighted fused lasso signal approximator.
\newblock {\em Preprint}.
\newblock arXiv:2407.18077.

\bibitem[Friedman et~al., 2008]{Friedman2008}
Friedman, J., Hastie, T., and Tibshirani, R. (2008).
\newblock Sparse inverse covariance estimation with the graphical lasso.
\newblock {\em Biostatistics}, 9:432--441.

\bibitem[Gibberd and Nelson, 2017]{gibberd2017regularized}
Gibberd, A.~J. and Nelson, J.~D. (2017).
\newblock {Regularized estimation of piecewise constant Gaussian graphical models: The group-fused graphical Lasso}.
\newblock {\em Journal of Computational and Graphical Statistics}, 26(3):623--634.

\bibitem[Grandt et~al., 2022]{Grandt2022}
Grandt, C.~L., Brackmann, L.~K., Poplawski, A., Schwarz, H., Hummel-Bartenschlager, W., Hankeln, T., Kraemer, C., Marini, F., Zahnreich, S., Schmitt, I., et~al. (2022).
\newblock {Radiation-response in primary fibroblasts of long-term survivors of childhood cancer with and without second primary neoplasms: The KiKme study}.
\newblock {\em Molecular Medicine}, 28(1):105.

\bibitem[Hallac et~al., 2017]{Hallac2017}
Hallac, D., Park, Y., Boyd, S., and Leskovec, J. (2017).
\newblock Network inference via the time-varying graphical lasso.
\newblock {\em Proceedings of the ACM SIGKDD International Conference on Knowledge Discovery and Data Mining}, Part F1296:205--213.

\bibitem[Hastie et~al., 2009]{hastie2009elements}
Hastie, T., Tibshirani, R., Friedman, J.~H., and Friedman, J.~H. (2009).
\newblock {\em {The elements of statistical learning: Data mining, inference, and prediction}}, volume~2.
\newblock Springer.

\bibitem[Lauritzen, 1996]{lauritzen1996}
Lauritzen, S. (1996).
\newblock {\em Graphical models}.
\newblock Oxford Statistical Science Series. Clarendon Press.

\bibitem[Liu and Wang, 2017]{Liu2017}
Liu, H. and Wang, L. (2017).
\newblock {TIGER: A tuning-insensitive approach for optimally estimating Gaussian graphical models}.
\newblock {\em Electronic Journal of Statistics}, 11:241--294.

\bibitem[Marron et~al., 2021]{Marron2021}
Marron, M., Brackmann, L.~K., Schwarz, H., Hummel-Bartenschlager, W., Zahnreich, S., Galetzka, D., Schmitt, I., Grad, C., Drees, P., Hopf, J., et~al. (2021).
\newblock Identification of genetic predispositions related to ionizing radiation in primary human skin fibroblasts from survivors of childhood and second primary cancer as well as cancer-free controls: Protocol for the nested case-control study kikme.
\newblock {\em JMIR Research Protocols}, 10(11):e32395.

\bibitem[Monti et~al., 2014]{Monti2014}
Monti, R.~P., Hellyer, P., Sharp, D., Leech, R., Anagnostopoulos, C., and Montana, G. (2014).
\newblock Estimating time-varying brain connectivity networks from functional {MRI} time series.
\newblock {\em Neuroimage}, 103:427--443.

\bibitem[Newman, 2018]{newman2018networks}
Newman, M. (2018).
\newblock {\em {Networks}}.
\newblock Oxford University Press.

\bibitem[P{\'o}sfai and Barabasi, 2016]{posfai2016network}
P{\'o}sfai, M. and Barabasi, A.-L. (2016).
\newblock {\em {Network science}}.
\newblock Citeseer.

\bibitem[Uhler, 2018]{Uhler2018}
Uhler, C. (2018).
\newblock Gaussian graphical models.
\newblock In {\em Handbook of graphical models}, pages 217--238. CRC Press.

\bibitem[Vinciotti et~al., 2024]{vinciotti2024random}
Vinciotti, V., Wit, E.~C., and Richter, F. (2024).
\newblock Random graphical model of microbiome interactions in related environments.
\newblock {\em Journal of Agricultural, Biological and Environmental Statistics}, pages 1--14.

\bibitem[Witten and Tibshirani, 2009]{Witten2009}
Witten, D.~M. and Tibshirani, R. (2009).
\newblock Covariance-regularized regression and classification for high dimensional problems.
\newblock {\em Journal of the Royal Statistical Society Series B: Statistical Methodology}, 71(3):615--636.

\bibitem[Wu et~al., 2019]{Wu2019}
Wu, N., Huang, J., Zhang, X.-F., Ou-Yang, L., He, S., Zhu, Z., and Xie, W. (2019).
\newblock Weighted fused pathway graphical lasso for joint estimation of multiple gene networks.
\newblock {\em Frontiers in Genetics}, 10:1--12.

\end{thebibliography}

\appendix

\section{Interpolation} \label{sec:appendix:interpolation}

As outlined in Section~\ref{sec:interpolation}, interpolating a graph $G_{m+1}$ using an estimated CVN model $\Tcvnest$ involves solving an optimization problem for each possible edge $\{s,t\}$ in the graph:
\begin{equation*}
    \wh{\theta}_{st}^{(m+1)} = \argmin_{\theta_{st}^{(m+1)} \in \mathds{R}} \lambda_1 |\theta_{st}^{(m+1)}| + \lambda_2 \normsimple{\bm{\omega}^\top \widehat{\bm{\theta}}_{st} - \theta_{st}^{(m+1)}}_1. 
\end{equation*}
Here, $\widehat{\theta}_{st}^{(m+1)}$ represents the interpolated entry for the edge $\{s,t\}$ in the precision matrix. The vector $\bm{\omega} = \left(\omega_1, \omega_2, \ldots, \omega_m \right)^\top$ contains predetermined, non-negative weights, and $\widehat{\bm{\theta}}_{st} = \left(\widehat{\theta}_{st}^{(1)}, \widehat{\theta}_{st}^{(2)}, \ldots, \widehat{\theta}_{st}^{(m)}\right)^\top$ are the estimated values in the precision matrices for edge $\{s,t\}$ of the $m$ graphs. The tuning parameters are $\lambda_1 > 0$ and $\lambda_2 \geq 0$.

Although this optimization problem lacks an analytical solution, it is convex, allowing for solving it numerically using derivative-free search methods. We employ Brent's algorithm for this purpose \citep{brent2013algorithms}. For ease of notation, let $x = \theta_{st}^{(m - 1)}$ and $\bm{y} = \widehat{\bm{\theta}}_{st}$. The goal is to solve:
\newcommand{\f}[1]{f(#1; \bm{y}, \bm{\omega})}
\begin{equation*}
    \argmin_{x \in \R} \f{x} = \argmin_{x \in \R} \lambda_1 |x| + \lambda_2 \norm{\bm{\omega}^\top \bm{y} - x}_1.
    \label{eq:optiminterpolation}
\end{equation*}
The convexity of the function $\f{x}$ in $x$ is evident since
\begin{equation*}
    \begin{split}
        \f{\alpha x_1 + (1 - \alpha)x_2} & \leq \alpha \f{x_1} + (1 - \alpha)\f{x_2}
    \end{split}
\end{equation*}
for all $x_1, x_2 \in \R$, $\alpha \in [0,1]$, $\lambda_1 > 0$, $\lambda_2 \geq 0$, $\bm{\omega} \in \mathds{R}_+^m$ and $\bm{y} \in \R^m$. This is easy to see for both 
$$
    | \alpha x_1 + (1 - \alpha) x_2| \leq \alpha |x_1| + (1 - \alpha) |x_2| 
$$
and 
$$
    \norm{\bm{\omega}^\top \bm{y} - (\alpha x_1 + (1 - \alpha) x_2)}_1 \leq \alpha \norm{\bm{\omega}^\top \bm{y} - x_1}_1 + (1 - \alpha) \norm{\bm{\omega}^\top \bm{y} - x_2}_1. 
$$
Note that the sign of $x$ (i.e., $\wh{\theta}_{st}^{(m+1)}$) is irrelevant in this context, as our objective is solely to determine the existence of the edge $\{s,t\}$ in the interpolated graph, i.e., whether $\wh{\theta}_{st}^{(m+1)} \neq 0$.

\end{document}